\documentclass{article}

\usepackage{amsmath}
\usepackage{graphicx}
\usepackage{subcaption}
\usepackage{setspace}
\usepackage{multirow}
\usepackage{caption}
\usepackage{longtable}
\usepackage[square]{natbib}
\usepackage{amssymb}
\usepackage{hyperref}
\usepackage[title]{appendix}
\usepackage[margin=1in]{geometry}
\usepackage[table,xcdraw]{xcolor}

\begin{document}
\bibliographystyle{plainnat}

\title{Well-balanced, anti-diffusive non-oscillatory central difference
(\textit{adNOC}) scheme for the shallow water equations with wet-dry fronts.}
\author{
Haseeb Zia
\thanks{
Dept. of Earth and Environmental Sciences,
Univ. of Geneva,
13 Rue des Maraichers,
1205 Geneva, Switzerland. 
E-mail: Haseeb.Zia@unige.ch}
\and Guy Simpson 
\thanks{
Dept. of Earth and Environmental Sciences,
Univ. of Geneva,
13 Rue des Maraichers,
1205 Geneva, Switzerland. 
E-mail: Guy.Simpson@unige.ch}
}

\maketitle
\doublespacing
\begin{abstract}
We present a central differencing scheme for the solution of the shallow water equations with non-flat bottom topography and moving wet-dry
fronts. The problem is numerically challenging due to two reasons. First, the non-flat bottom topography requires accurate balancing of the source term of the momentum conservation equation accounting for the gravitational force and the flux gradient term accounting for the force due to pressure imbalance. Second, the modelling of moving wet-dry fronts involves handling of diminishing water height, which is numerically challenging to handle. The Riemann-solver free scheme is fast, simple and robust. It successfully avoids negative water
depths at moving wet-dry boundaries and it exhibits good balancing between flux gradients and  source terms. The performance of the scheme is verified with a number
of test cases and the results compare favorably with published analytical solutions.
\end{abstract}

\section{Introduction}
The shallow water equations are a set of hyperbolic conservation laws obtained by
depth averaging of the Navier-Stokes equations that are widely used to model free surface water flows. In one dimension,
the shallow water equations  can be written as:
\begin{equation}
\frac{\partial h}{\partial t}+\frac{\partial q}{\partial x}=0,\label{eq:1}
\end{equation}

\begin{equation}
\frac{\partial q}{\partial t}+\frac{\partial}{\partial
x}\left(qu+\frac{1}{2}gh^2\right)=-gh\frac{\partial z}{\partial
x}-ghS_{f},\label{eq:2}
\end{equation}
where $h$ is the water depth, $q$ is the flux, $u$ is the horizontal velocity, $g$ is acceration due to
gravity, $z$ is the elevation of the bottom topography, $S_f$ is the friction slope, for wich we have used
the widely used Manning's formulation given by $S_{f}=\frac{n^{2}u|u|}{h^{4/3}}$ and $n$ is  Manning's roughness coefficient.
Numerical solution of the shallow water equations is challenging and has been the subject of numerous studies over the last few decades
(e.g.
\cite{leveque1998balancing,hubbard2000flux,rogers2003mathematical,castro2005numerical,audusse2004fast,kurganov2007second,bryson2011well,xing2010positivity,Bollermann2012,bollermann2015finite}).
However, two problems have demanded special attention and continue to drive the development of  new and improved schemes.

The first problem is the the difficulty of accurately calculating the steady state solution  for a  water body at rest over an irregular bottom topography. In this case,  a delicate balance is required between the  flux gradient and the source term in the
momentum equation. Even a small imbalance between the terms results in unphysical spontaneous water movement that may completely destroy the solution.
 Since the early works of LeRoux et al. \cite{greenberg1996well} and
Bermudez et al. \cite{bermudez1994upwind}, schemes capable of exactly balancing
the two terms have been called well-balanced schemes or schemes upholding the 
C-property. Some examples of successful well balanced schemes are published by 
\cite{leveque1998balancing,hubbard2000flux,rogers2003mathematical,castro2005numerical,audusse2004fast,tseng2004improved,audusse2005well,xing2005high,xing2006new,marche2007evaluation,kurganov2007second,Canestrelli2010291,kesserwani2010well,ricchiuto2011c,Bollermann2012,bollermann2015finite}.

The second unrelated problem concerns the computation of the solution when the water
depth $h$ diminishes to a very small value or zero. This problem arises often when the boundary
between the wet and dry areas of the domain is mobile, a situation commonly
encountered in nature, e.g.,  storms,  tsunamis, river bars,
breaking of dams. This problem is
especially important for schemes that use eigenvalues of the Jacobian of the
flux function to calculate the solution; because the eigenvalues are given
by $u\pm \sqrt{gh}$, the calculation would  break down if $h$ becomes
negative. Thus,  it is essential that these numerical  schemes  preserve positivity of
the water depth. This condition becomes increasingly difficult to satisfy when the water depth is
close to zero because even small numerical oscillations can cause negative values. Examples of some 
positivity preserving, well balanced schemes are published by
\cite{audusse2004fast,audusse2005well,kurganov2007second,bryson2011well,xing2010positivity}.

In this paper we present a fast and robust  numerical  scheme which
adequately meets both  challenges; i.e.,  it is well-balanced and preserves positivity.  The scheme is based on the centred approach initially presented by Lax and Friedrich in \cite{friedrichs1971systems} and later developed
further by Nessyahy and Tadmor \cite{NessyahuTadmor1990}. This approach provides several advantages over Godunov type schemes that rely on Riemann-solvers and eigenvalues to calculate fluxes. In the case of shallow water equations, the
central approach holds special advantage since it avoids calculation of the eigenvalues  (necessary for Riemann-based)  and  therefore also
avoids break down of calculations in the case of negative water depth. Although positivity preservation is important for realistic results,
it is not essential for the stability of the  scheme. While some other Riemann-free central schemes published in literature, such as the
central-upwind schemes (\cite{kurganov2002central,kurganov2007second,bryson2011well,Bollermann2012,bollermann2013well}) and \textit{PRICE}
schemes \cite{canestrelli2009well,canestrelli2010well} still rely on partial knowledge of the eigenvalues and the largest eigenvalues are
still calculated, our scheme involves no calculation or knowledge of eigenvalues to calculate fluxes. The presented scheme (referred to as
\textit{adNOC}) is very simple and universal and is particularly suitable for models where the shallow water equations are coupled with other models, e.g.
pollutant or sediment transport models.

This manuscript  is structured as follows. In section \ref{sec:adnoc}, a brief
description of the \textit{adNOC} scheme is presented. Section \ref{sec:well},
describes how the flux gradient and the source terms are balanced. In
section \ref{sec:well-ballanced adnoc}, a well-balanced \textit{adNOC}
scheme is presented for the shallow water equations. Section
\ref{sec:Positivity} discusses the positivity preservation of the scheme and
finally section \ref{sec:results} presents  numerical results and 
comparison with benchmarks and  analytical solutions.

\section{The \textit{adNOC} Scheme}
\label{sec:adnoc}
This work is based on the  \textit{adNOC} numerical scheme
introduced in \cite{Zia2015adNOC}. This method is an   extension of the Nessyahu-Tadmor non-oscillatory central (NOC)
scheme, modified to reduce numerical dissipation that is especially severe in the classic NOC scheme under certain situations. The NOC scheme is a second-order method that utilizes the mid-point
rule to achieve second order accuracy in both space and time. The method
operates in predictor-corrector fashion where the solution is first
approximated at half time step in the predictor step which is then used in the
corrector step to realise the second order solution.

Central differencing schemes  are similar to Godunov type methods in the sense that they operate by evaluating cell averages through
piece-wise cell reconstructions. The difference between the two approaches is the use of
staggered cells for the evaluation of these averages in central schemes.
This use of staggered grids allows central schemes to calculate fluxes straight
forwardly as the reconstructed solutions are smooth at the points where fluxes are being calculated.
Upwind based Godunov schemes on the other hand, use Riemann-solvers to calculate
fluxes which are complicated and mostly problem specific. However,
the drawback of the use of staggered grids by the central schemes is the
numerical dissipation which causes the solution to smear, resulting in
inaccurate results. The \textit{adNOC} scheme reduces this diffusion by adding an
anti-diffusion correction to the \textit{NOC} scheme \cite{Zia2015adNOC}.

Consider the following scalar hyperbolic conservation law
\begin{equation}\label{eq:3}
\frac{\partial w}{\partial t}+\frac{\partial f(w)}{\partial
x}=s(w),
\end{equation}
where $w$ is the conserved quantity, $f$ is the flux and $s$ is
the source term,
both functions of $w$. Central schemes proceed by discretizing this equation as 
 \begin{equation}\label{eq:4}
\overline{w}_{j+1/2}^{n+1}=\frac{1}{2}
(\overline{w}_{j+1/4}^{n}+\overline{w}_{j+3/4}^{n})-\frac{\Delta
t}{\Delta x}(f_{j+1}^{n+1/2}-f_{j}^{n+1/2})+\frac{\Delta
t}{2}(s_{j+1/4}^{n+1/2}+s_{j+3/4}^{n+1/2}),
\end{equation}
where  $\overline{w}^n_j$ is the cell average at the $n^{th}$ time step, $\Delta x$ is the constant  grid spacing  and $j$  is the space index. The terms $\overline{w}_{j+1/4}^{n}$ and $\overline{w}_{j+3/4}^{n}$
can be approximated  using the Taylor series reconstructions
\begin{equation}\label{eq:5}
\overline{w}_{j+1/4}^{n}=\overline{w}_{j}^{n}+\frac{\Delta
x}
{4}\sigma(w)_{j}^{n}\;\; \text{and}
\;\;\overline{w}_{j+3/4}^{n}=\overline{w}_{j+1}^ {n}-\frac{\Delta
x}{4}\sigma(w)_{j+1}^{n},
\end{equation}
where $\sigma(w)$ is the discrete spatial slope of the variable $w$. The fluxes
in Eq.
(\ref{eq:4}) are approximated by:
\begin{equation}\label{eq:6}
f_{j}^{n+1/2}=f(\overline{w}_{j}^{n+1/2})\;\;\text{and}\;\;
f_{j+1}^{n+1/2}=f(\overline{w}_{j+1}^{n+1/2}).
\end{equation}
Similarly, the  source terms in Eq. (\ref{eq:4}) can be  approximated by:
\begin{equation}\label{eq:7}
s_{j+1/4}^{n+1/2}=s(\overline{w}_{j+1/4}^{n+1/2})\;\;\text{and}\;\;
s_{j+3/4}^{n+1/2}=s(\overline{w}_{j+3/4}^{n+1/2}),
\end{equation}
where
\[
\overline{w}_{j+1/4}^{n+1/2}=\overline{w}_{j}^{n+1/2}+\frac{\Delta
x}
{4}\sigma(w)_{j}^{n}\;\;\text{and}\;\;\overline{w}_{j+3/4}^{n+1/2}=\overline{w}_{j
+1}^{n+1/2}-\frac{\Delta
x}{4}\sigma(w)_{j+1}^{n}.
\]
Notice that the calculation of the flux and the source terms requires solution
at the half time step, which is calculated using the conservation law in the predictor step:
\begin{equation}\label{eq:8}
\overline{w}_{j}^{n+1/2}=\overline{w}_{j}^{n}-\frac{\Delta
t}{2}(\sigma(f)_{j})^{n}+\frac{\Delta t}
{2}s(\overline{w}_{j}^{n}),
\end{equation}
where $\sigma(f)$ is the discrete spatial slope of the
flux.

The second order accuracy of the scheme resolves shocks and
discontinuities in the solution at the expense of causing spurious
oscillations. To ensure the stability of the scheme, NOC methods use the concept of TVD (total variation diminishing),
which ensures that no oscillations originate in the vicinity of shocks in
the solution. To fulfill the TVD condition, the slopes $\sigma(w)$ and
$\sigma(f(w))$ should be evaluated with limiters (proof can be seen in
\cite{NessyahuTadmor1990}). A large selection of limiters are available  (see e.g. \cite{sweby1984high,leveque2002finite}), any of which
can be used with the scheme. In our work, we have used the standard \textit{minmod}
limiter given by:
\[
minmod\{r_1,r_2\}=\frac{1}{2}[sgn(r_1)+sgn(r_2)].Min(|r_1|,|r_2|)
\]
where $r_1$ and $r_2$ are the slopes at successive positions on the solution
mesh.

The high resolution Nessyahu-Tadmor scheme presented above
performs well in resolving shocks and discontinuities. However, in certain
cases, the scheme becomes impractical to use due to excessive numerical dissipation.
Central schemes are known to be highly diffusive especially when small time steps are used \cite{kurganov2000new,huynh2003analysis,kurganov2007reduction,siviglia2013numerical,canestrelli2012restoration}.
The reason for this numerical dissipation can be observed directly from the
scheme (Eq. \ref{eq:4}). Notice that if a small time step is used or the system is
close to steady state, the second and third terms on the right hand side of the
equation becomes negligible and the scheme is reduced to an averaging function
i.e., the first term. To eliminate the numerical dissipation caused by this
averaging, an anti-diffusion correction can be added to the \textit{NOC} scheme,
yielding the \textit{adNOC} scheme. The correction is based on correct
evaluation of the slopes for reconstructions in Eq.
(\ref{eq:4}) (see \cite{Zia2015adNOC} for details). The \textit{NOC} scheme
with the anti-diffusion correction (\textit{adNOC} scheme) is given by:
\[
\overline{w}_{j+1/2}^{n+1}=\frac{1}{2}
(\hat{w}_{j+1}^{n}+\hat{w}_{j}^{n})+\frac{\Delta x}{8}(1-
\varepsilon)(\sigma_{j}^{n}-\sigma_{j+1}^{n})-\frac{\varepsilon}
{4}(\overline{w}_{j+3/2}^{n-1}-2\overline{w}_{j+1/2}^{n-
1}+\overline{w}_{j-1/2}^{n-1})
\]
\begin{equation}\label{eq:9}
\;\;\;\;-\lambda(f_{j+1}^{n+1/2}-f_{j}^{n+1/2})+\frac{\Delta
t}{2}(s_{j+1/4}^{n+1/2}+s_{j+3/4}^{n+1/2}).
\end{equation}
where $\hat{w}$ is the cell average evaluated without the anti-diffusion
correction i.e. the third term in the right hand side of (\ref{eq:9}). The
parameter $\varepsilon$ is the variable specifying the strength of the
corrected slope against the limited slope. A value of 1 for
$\varepsilon$ signifies that only corrected slopes are used for the
reconstructions (i.e. fully \textit{adNOC}
scheme) while a value of 0 signifies the standard limited slope evaluations
are used, which reduces Eq. (\ref{eq:9}) to Eq. (\ref{eq:4}) (i.e. standard
\textit{NOC} scheme). This splitting of slopes is necessary for the stability of
the scheme (see \cite{Zia2015adNOC} for details).

\section{Balancing}
\label{sec:well}
A numerical scheme solving the system of shallow water equations is said to be
well-balanced, or maintaining the C-property (conservation property),  if the momentum
flux and momentum source terms are balanced for nearly hydrostatic flows
where $u\ll \sqrt{gh}$. The terms to be balanced are the pressure driven part of
the momentum flux and the source term due to bed slope:
\[
\frac{\partial}{\partial x}\left(\frac{gh^{2}}{2}\right)=-gh\frac{\partial z}{\partial x}
\]
The balancing of these terms becomes especially important in quasi steady state
scenarios. For example, if the two terms are not balanced in the case of a lake
at rest, the momentum equation produces artificial momentum, which results in a so called
"numerical storm''. In some cases with course grids, the waves produced by the
imbalance can be higher in magnitude than those of the true
solution. The error in the balancing of the flux and the source terms is
related to discretization error (i.e., erroneous evaluation
of the partial differentials). To see this, consider the following simple one dimensional central difference
discretization of the flux and  source terms:
\begin{equation}\label{eq:10}
\frac{\partial}{\partial x}\left(\frac{gh^{2}}{2}\right)=\frac{g}{2}\left(\frac{h_{i+1/2}^{2}-h_{i-1/2}^{2}}{\Delta x}\right)
=\frac{g}{2}\left(\frac{h_{i+1/2}-h_{i-1/2}}{\Delta x}\right)\left(h_{i+1/2}+h_{i-1/2}\right)
\end{equation}
and
\begin{equation}\label{eq:11}
-gh\frac{\partial z}{\partial
x}=-gh_i\left(\frac{z_{i+1/2}-z_{i-1/2}}{\Delta x}\right).
\end{equation}
A 'lake at rest' scenario  is given by the conditions:
\[
h+z=H=constant\;\;\;\text{and}\;\;\; u=0,
\]
where $H$ is the water surface height which is a constant over the wet part of
the domain. By taking the derivative of the condition, it can be
seen that the differentials of water depth $h$ and bed height $z$ are equal in
magnitude but  opposite in sign:
\[
\frac{\partial h}{\partial x}=-\frac{\partial z}{\partial x}.
\]
Note from Eq. (\ref{eq:10}) and Eq. (\ref{eq:11}) that, in the case of a lake at
rest, the two terms do not balance completely but approach each other as the grid spacing is
decreased, i.e., if we take the limit $\lim_{\Delta x \to 0}$, the two equations converge.
\[
\lim_{\Delta x\to0}\frac{g}{2}\left(\frac{h_{i+1/2}-h_{i-1/2}}{\Delta x}\right)\left(h_{i+1/2}+h_{i-1/2}\right)=-gh_{i}\frac{\partial z}{\partial x}
\]
To remove this discretization error, we will first use the chain rule to obtain
the differential of the gravity driven flux term and then discretize it in the
same manner as the source term so that both suffer from same
discretization error, hence balancing each other. By
using the chain rule:
\[
\frac{\partial}{\partial x}\left(\frac{gh^{2}}{2}\right)=gh\frac{\partial h}{\partial x}
\]
which is exactly balanced by the source term if the same finite difference
operator is used for calculating both differentials.

\section{Well-balanced \textit{adNOC} scheme}
\label{sec:well-ballanced adnoc}
A well-balanced \textit{adNOC} scheme arises straightforwardly from the
\textit{adNOC} scheme. For the continuity equation \ref{eq:1}, a well balanced approximation is:
\begin{equation}\label{eq:12}
h_{j+1/2}^{n+1}=\frac{1}{2}
(\hat{h}_{j+1}^{n}+\hat{h}_{j}^{n})+\frac{\Delta x}{8}(1-
\varepsilon)(\sigma(h)_{j}^{n}-\sigma(h)_{j+1}^{n})-\frac{\varepsilon}
{4}(h_{j+3/2}^{n-1}-2h_{j+1/2}^{n-
1}+h_{j-1/2}^{n-1})
-\lambda(q_{j+1}^{n+1/2}-q_{j}^{n+1/2})
\end{equation}
where $\sigma(h)$ is the limited slopes evaluated using the \textit{minmod} limiter
and $q^{n+1/2}$ is the flux at the  half time step evaluated using the conservation law
(see Eq. \ref{eq:8}).
\[
q_{j}^{n+1/2}=q_{j}^{n}-\frac{\Delta
t}{2}(\sigma(f(q))_{j}^{n}+\frac{\Delta t}
{2}S(q_{j}^{n}),
\]
where $S(q)$ is the source function for the momentum conservation equation (i.e.
the right hand side of Eq. \ref{eq:13}). It is important to mention here that the
flux and source function should be balanced as discussed in the previous
section. For the momentum conservation equation (\ref{eq:2}),  we will rearrange
it by splitting the flux gradient term into the convective acceleration term and the pressure gradient term.
The momentum equation can be rewritten by moving the pressure gradient term to
the right hand side:
\begin{equation}
\frac{\partial q}{\partial t}+\frac{\delta}{\partial
x}(qu)=-gh\left(\frac{\partial h}{\partial x}+\frac{\partial z}{\partial
x}\right)-ghS_f.\label{eq:13}
\end{equation}
The solution of this equation for the flux $q$ is:
 \[
q_{j+1/2}^{n+1}=\frac{1}{2}
(\hat{q}_{j+1}^{n}+\hat{q}_{j}^{n})+\frac{\Delta x}{8}(1-
\varepsilon)(\sigma(q)_{j}^{n}-\sigma(q)_{j+1}^{n})-\frac{\varepsilon}
{4}(q_{j+3/2}^{n-1}-2q_{j+1/2}^{n-
1}+q_{j-1/2}^{n-1})
\]
\begin{equation}
\;\;\;\;-\lambda\left(\frac{(q_{j+1}^{n+1/2})^2}{h_{j+1}^{n+1/2}}-\frac{(q_{j}^{n+1/2})^2}{h_{j}^{n+1/2}}\right)+\frac{\Delta
t}{2}\left(S(q_{j+1/4}^{n+1/2})+S(q_{j+3/4}^{n+1/2})\right).\label{eq:14}
\end{equation}
where $S(q)$ is the right hand side of Eq. (\ref{eq:13}) calculated by:
\begin{equation}\label{eq:15}
S(q_{j+1/4}^{n+1/2})=-gh_{j+1/4}^{n+1/2}\left(\left(\frac{\partial h}{\partial x}\right)_{j+1/4}^{n+1/2}+\left(\frac{\partial z}{\partial x}\right)_{j+1/4}\right)-gh_{j+1/4}^{n+1/2}S_{f}
\end{equation}
\begin{equation}\label{eq:16}
S(q_{j+3/4}^{n+1/2})=-gh_{j+3/4}^{n+1/2}\left(\left(\frac{\partial h}{\partial x}\right)_{j+3/4}^{n+1/2}+\left(\frac{\partial z}{\partial x}\right)_{j+3/4}\right)-gh_{j+3/4}^{n+1/2}S_{f}
\end{equation}
To calculate slopes, we  use central difference approximations:
\[
\left(\frac{\partial h}{\partial
x}\right)_{j+1/4}^{n+1/2}=\frac{h_{j+3/4}^{n+1/2}-h_{j-1/4}^{n+1/2}}{\Delta
x},\;\;\;\left(\frac{\partial h}{\partial x}\right)_{j+3/4}^{n+1/2}=\frac{h_{j+5/4}^{n+1/2}-h_{j+1/4}^{n+1/2}}{\Delta x} 
\]
and
\[
\left(\frac{\partial z}{\partial
x}\right)_{j+1/4}=\frac{z_{j+3/4}-z_{j-1/4}}{\Delta
x},\;\;\;\left(\frac{\partial z}{\partial
x}\right)_{j+3/4}=\frac{z_{j+5/4}-z_{j+1/4}}{\Delta x}.
\]
 \subsection*{Verification of the C-property}
We  now show that the above \textit{adNOC} scheme exactly maintains the 'lake atrest' condition. Consider Eq. (\ref{eq:12}), which is the
discrete form of the  continuity equation. Since we assume the lake is at rest initialy, the flux terms are equal to zero. Here, we use
full anti-diffusive slopes i.e. $\varepsilon=1$. The solution at any cell $j$ and time step $n+1$ with the \textit{adNOC} scheme is given by:
\begin{equation}
h_{j}^{n+1}=\frac{1}{2}
(\hat{h}_{j+1/2}^{n}+\hat{h}_{j-1/2}^{n})-\frac{1}
{4}(h_{j+1}^{n-1}-2h_{j}^{n-
1}+h_{j-1}^{n-1}).\label{eq:17}
\end{equation}
Substituting
\[
\hat{h}_{j+1/2}^{n}=h_{j+1}^{n-1}+h_{j}^{n-1}\;\;\;\text{and}\;\;\;\hat{h}_{j-1/2}^{n}=h_{j}^{n-1}+h_{j-1}^{n-1}
\]
into Eq. (\ref{eq:17}) reduces it to:
\[
h_{j}^{n+1}=h_{j}^{n-1}
\]
showing that the solution is exactly maintained. To show that no artificial momentum is generated, we check weather the flux remain zero in
the case of lake at rest. We see that all the terms in \ref{eq:14}, except the source term, involves flux or flux gradients which are
zero. For the source term, we see that the friction slope $S_f$ is zero. Also From Eq. (\ref{eq:8}) for water depth $h$, we see that
$h^{n+1/2}_j=h^n_j$. Thus, Eq. (\ref{eq:15} and \ref{eq:16}) can be written as:
\[
S(q_{j+1/4}^{n+1/2})=-gh_{j+1/4}^{n}\left(\frac{H_{j+3/4}^{n}-H_{j-1/4}^{n}}{\Delta x}\right)
\]
and
\[
S(q_{j+3/4}^{n+1/2})=-gh_{j+3/4}^{n}\left(\frac{H_{j+5/4}^{n}-H_{j+1/4}^{n}}{\Delta x}\right)
\]
both of which are zero since $H$ is constant.

\subsection*{Partially wet cells}
\label{sec:partially}
\begin{figure}
	\centering
	\includegraphics{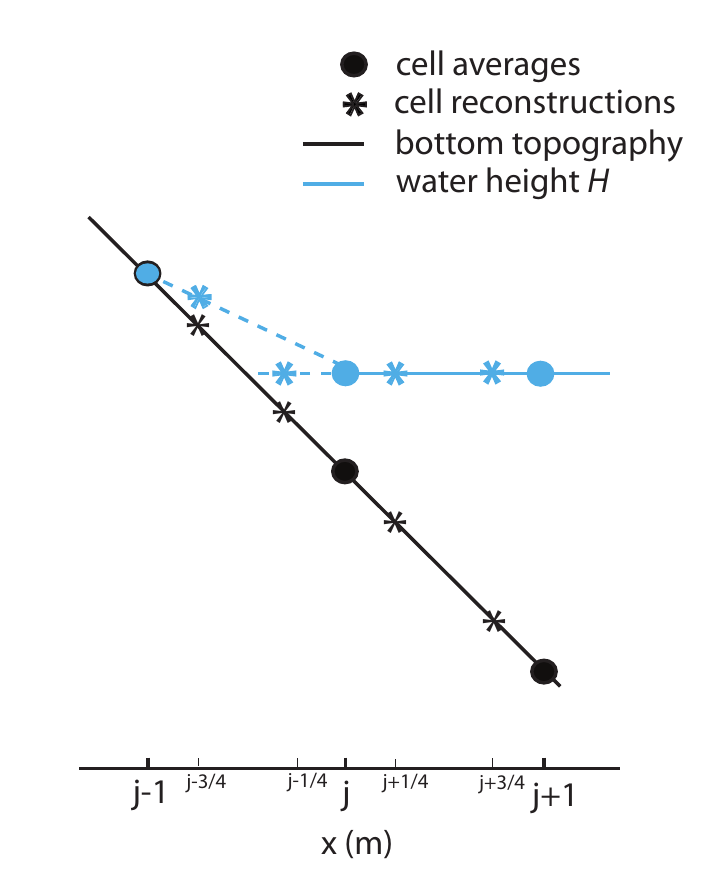}
		\caption{Cell reconstructions in the case of a partially wet cell.  }
	\label{fig:pwet}
\end{figure}

The previous section shows that the proposed \textit{adNOC} scheme is well balanced when cells are fully wet. However, in the case of 
partially wet cells, maintaining the C-property is not  trivial. As
discussed earlier, the differentials of water depth and bed morphology are equal
in magnitude with opposite sign in the case of a  lake at rest. For partially wet
cells, the balance does not hold unless the cell-center of the discretized bed aligns exactly 
with the water surface height. Figure \ref{fig:pwet} shows an example when the
water height $H$ calculated using  cell reconstructions.
It can be seen that the slopes will be balanced in the fully wet cell
$C_{j+1/2}$, but not in the partially wet cell $C_{j-1/2}$. This will cause 
artificial momentum and in turn artificial velocities at the wet/dry boundaries, resulting 
in a numerical storm. To avoid this situation, the source and flux terms are
forcibly balanced by equating the water depth slope to the bed topography slope. Thus,  for
a cell $C_{j+1/2}$ at a wet/dry boundary (i.e. $if\;\;h_j=0\;\cdot\; h_{j+1}>0\;\cdot\;(h_{j+1}+z_{j+1}\leq h_{j}+z_{j})\;\;|\;\;if\;\;
h_j+1=0\;\cdot\; h_{j}>0\;\cdot\;(h_{j}+z_{j}\leq h_{j+1}+z_{j+1})$, where ``$\cdot$'' signifies logical ``and'', and ``$|$''
signifies logical ``or''):
\[
 \begin{cases} \sigma(h_{j+3/4})=\sigma(z_{j+3/4})
 \\
\sigma(h_{j+1/4})=\sigma(z_{j+1/4})
 \end{cases}
\]

\section{Positivity-preservation}
\label{sec:Positivity}
Positivity preservation is the second desirable property for a scheme solving
the shallow water equations with wet-dry fronts. This property requires that the
water depth remains positive to ensure stable operation of the scheme.
Preserving Positivity of water depth also makes physical sense since a negative water depth
is clearly unrealistic. As discussed before, this property is more important for 
schemes which use Riemann-solvers and/or eigenvalues to calculate the solution (e.g.
\cite{leveque1998balancing,castro2005numerical,bryson2011well,bollermann2013well,berthon2008positive,FLD:FLD285,canestrelli2009well,chertock2015well,vcrnjaric2004balanced,hubbard2000flux,kurganov2007second}).
Some other schemes solve for the water level $H$ instead of water
depth $h$ (e.g. \cite{bryson2011well,liang2009adaptive,Bollermann2012}). For
these schemes, a water level of less than the bed level would result in a negative water depth since the water depth is calculated by
subtracting the bed height $z$ from the water level $H$. This would also lead to complex eigenvalues and
breakdown of Riemann-solvers. For the vast majority of  schemes published in
literature, a negative water depth can cause breakdown of calculation or stability problems.
For the \textit{adNOC} scheme, since no eigenvalues  are utilized, no breakdown occurs. The condition,
nevertheless, is important for realistic results and for other functions
dependent on $h$, such as the friction slope. Also, since the
velocity $u$ is calculated by dividing the flux $q$ with the water
depth $h$, negative water depth will reverse the direction of the
velocity and will cause stability problems. To calculate velocity,
we are use the desingularization formula used by Kurganov et al.
\cite{kurganov2007second,chertock2015well}.
\[
u=\frac{2h(q)}{h^{2}+max(h^{2},\theta)}
\]
where $\theta$ is a small constant (on the order of $10^{-6}$).

The  \textit{adNOC} scheme may generate 
negative  water depths if the calculated flux takes out more water than what is present in the cell, due to an excessively large time step.
To ensure positivity, the time step should be limited so that this does not happen.
Consider the solution of Eq. (\ref{eq:1}) of the shallow water equations with
\textit{adNOC} scheme ($\varepsilon=1$):
\[
h_{i}^{n+1}=h_{i}^{n-1}-\frac{\Delta t^{n}}{\Delta x}\left(q{}_{i+1/2}^{n+1/2}-q{}_{i-1/2}^{n+1/2}\right)-\frac{\Delta t^{n-1}}{2\Delta x}\left(q{}_{i+1}^{n-1/2}-q{}_{i-1}^{n-1/2}\right)
\]
For $h_i$ to remain positive at time step $n+1$, the time step $\Delta t$ should
be restricted so that:
\[
\frac{\Delta t^{n}}{\Delta x}\left(q{}_{i+1/2}^{n+1/2}-q{}_{i-1/2}^{n+1/2}\right)+\frac{\Delta t^{n-1}}{2\Delta x}\left(q{}_{i+1}^{n-1/2}-q{}_{i-1}^{n-1/2}\right)\leq h_{i}^{n-1}
\]
\[
Cn(h_{i+1/2}^{n+1/2}-h_{i+1/2}^{n+1/2})+\frac{Cn}{2}(h_{i+1}^{n-1/2}-h_{i-1}^{n-1/2})\leq h_{i}^{n-1}
\]
where $C_n$ is the courant number given by $C_n=\frac{\Delta t\;max(u)}{\Delta
x}$. We see that, for the scheme to be positivity preserving, the Courant number
used to calculate the time step $\Delta t$ has to fulfill the following
condition:
\[
Cn\leq\frac{h_{i}^{n-1}}{h_{i+1/2}^{n+1/2}-h_{i+1/2}^{n+1/2}+\frac{1}{2}(h_{i+1}^{n-1/2}-h_{i-1}^{n-1/2})}
\]

\section{Numerical results}
\label{sec:results}
The \textit{adNOC} scheme has been tested with a number of benchmark tests and
the results are compared with the analytical solutions.

\subsection{One-dimensional quiescent flow over a bump}
This benchmark was proposed by Goutal and Maurel \cite{goutal1997proceedings} in
a dam-break wave simulation workshop held in 1997. Since then, it has been used by
many studies to validate the correct evaluation of the source and the
ability of the numerical scheme to converge in time towards steady flow
(e.g. see
\cite{Crnjaric-Zic:2004,Canestrelli2010291,audusse2004fast,berthon2008positive,liang2009adaptive,marche2007evaluation,rogers2003mathematical,valiani2006divergence,xing2006new}).
The experimental setup consists of a one-dimensional frictionless channel with
the length of 25 m. The bottom topography is given by:

\[
 z(x)=\begin{cases} 0.2-0.05(x-10)^2\;m  &\mbox{if } 8\;m\leq x\leq12\;m  \\
0 &\mbox{ }\text{otherwise} \end{cases}
\]
The test case is started with the initial condition of horizontal
free surface profile and the solution is evolved in time until a
steady state is reached. The initial conditions are as follows:
\[
q(x,0)=0\;\;\;\text{and}\;\;\;h+z=5\;m.
\]
The mode of flow (subcritical, transcritical or supercritical) in this
experiment depends on the water height $H$ and the discharge $q$ at the
upstream and downstream boundaries. We have tested all three flow modes in three
different test cases, the analytical solutions of which are provided by Goutal
and Maurel \cite{goutal1997proceedings}. The boundary conditions for the three
test cases are shown in Table \ref{table:tab}. Test case (1) is set up to get a
completely subcritical steady state flow, test case (2) is set up to obtain a
transcritical flow steady state with a shock, while test case (3) is set up to
obtain a transcritical flow steady state without a shock.
\begin{table}[]
\centering
\begin{tabular}{ccc}
\hline
\rowcolor[HTML]{EFEFEF} 
test case & q($x=0$) ($m^2/s$) & H($x=25 m$) ($m$) \\ \hline
\rowcolor[HTML]{FFFFFF} 
(1) & 4.42 & 2.0 \\
\rowcolor[HTML]{FFFFFF} 
(2) & 0.18 & 0.33 \\
\rowcolor[HTML]{FFFFFF} 
(3) & 1.53 & 0.66 \\ \hline
\end{tabular}
\caption{Boundary conditions for the one-dimensional quiescent flow over a bump
test cases.}
\label{table:tab}
\end{table}

\begin{figure}
\centering
\begin{subfigure}{.5\textwidth}
  \centering
  \includegraphics[width=\linewidth]{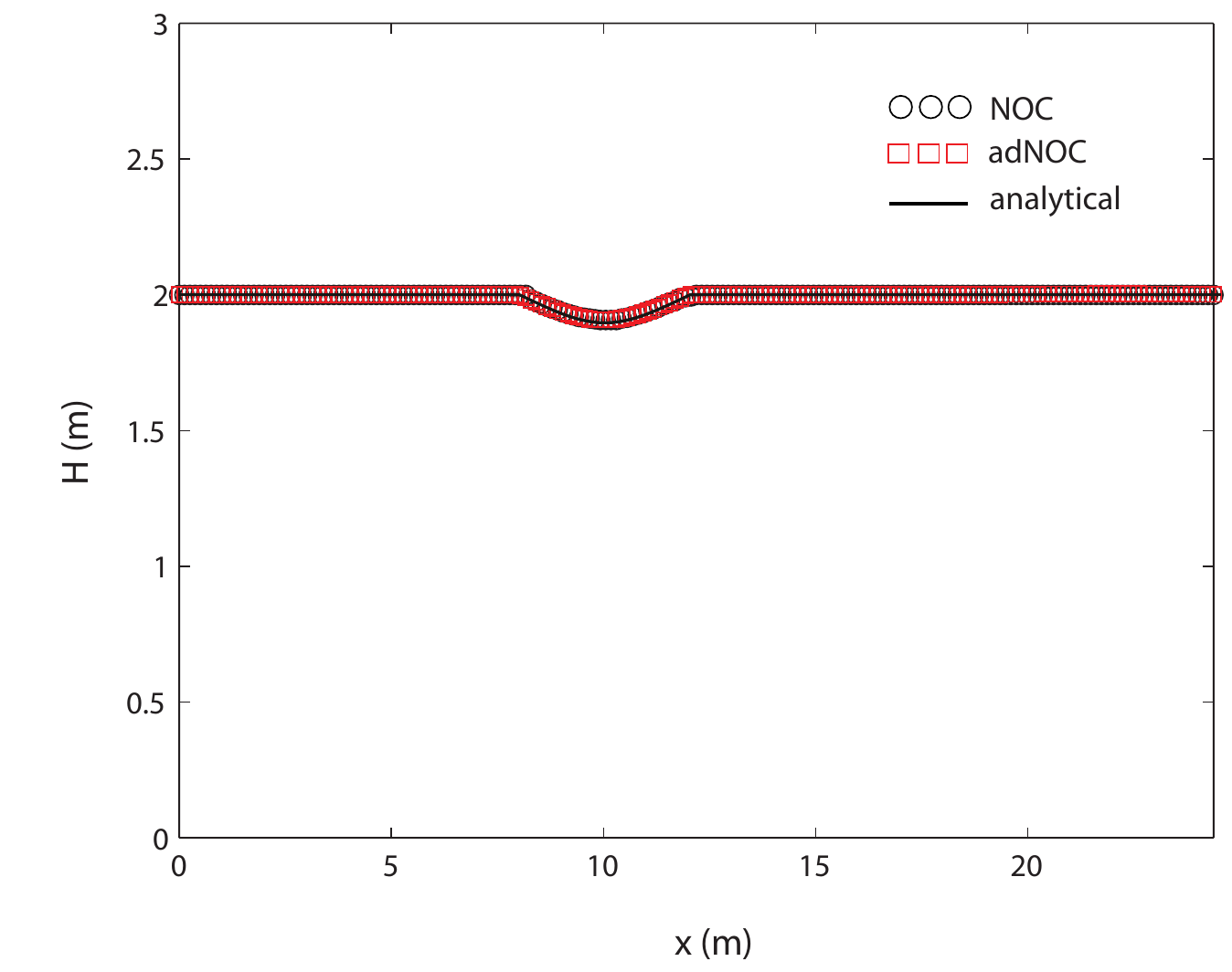}
\end{subfigure}%
\begin{subfigure}{.5\textwidth}
  \centering
  \includegraphics[width=\linewidth]{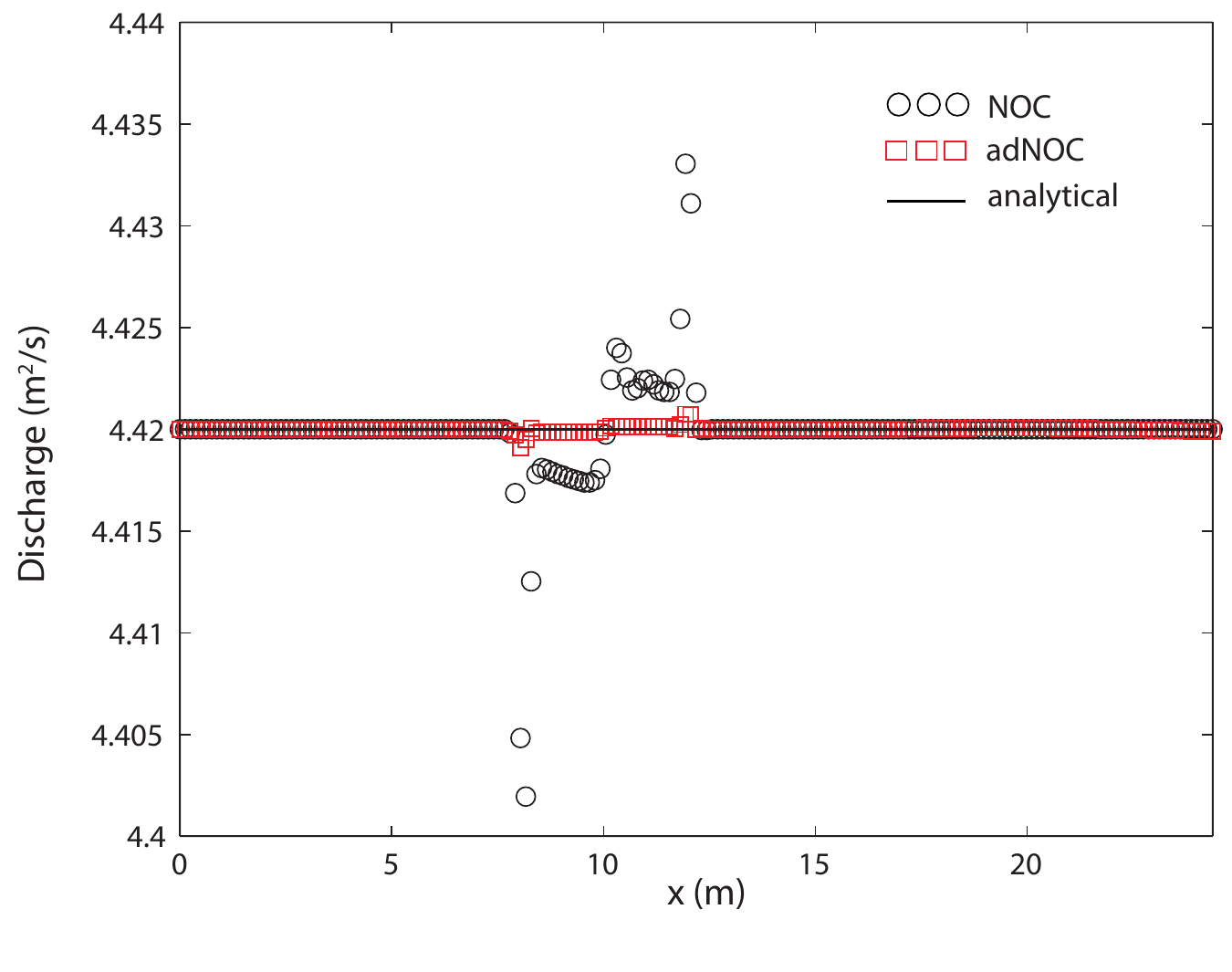}
\end{subfigure}

\begin{subfigure}{.5\textwidth}
  \centering
  \includegraphics[width=\linewidth]{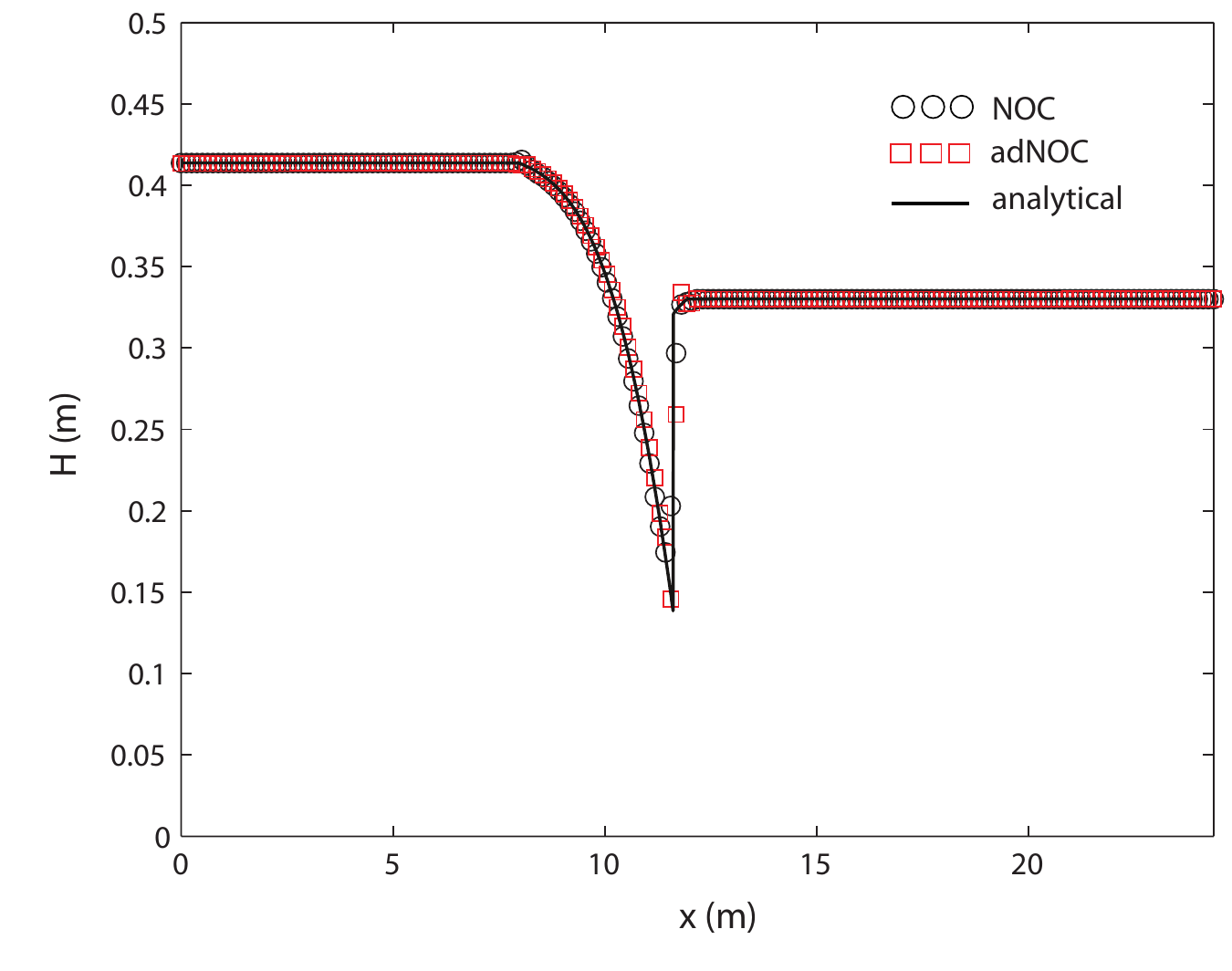}
\end{subfigure}%
\begin{subfigure}{.5\textwidth}
  \centering
  \includegraphics[width=\linewidth]{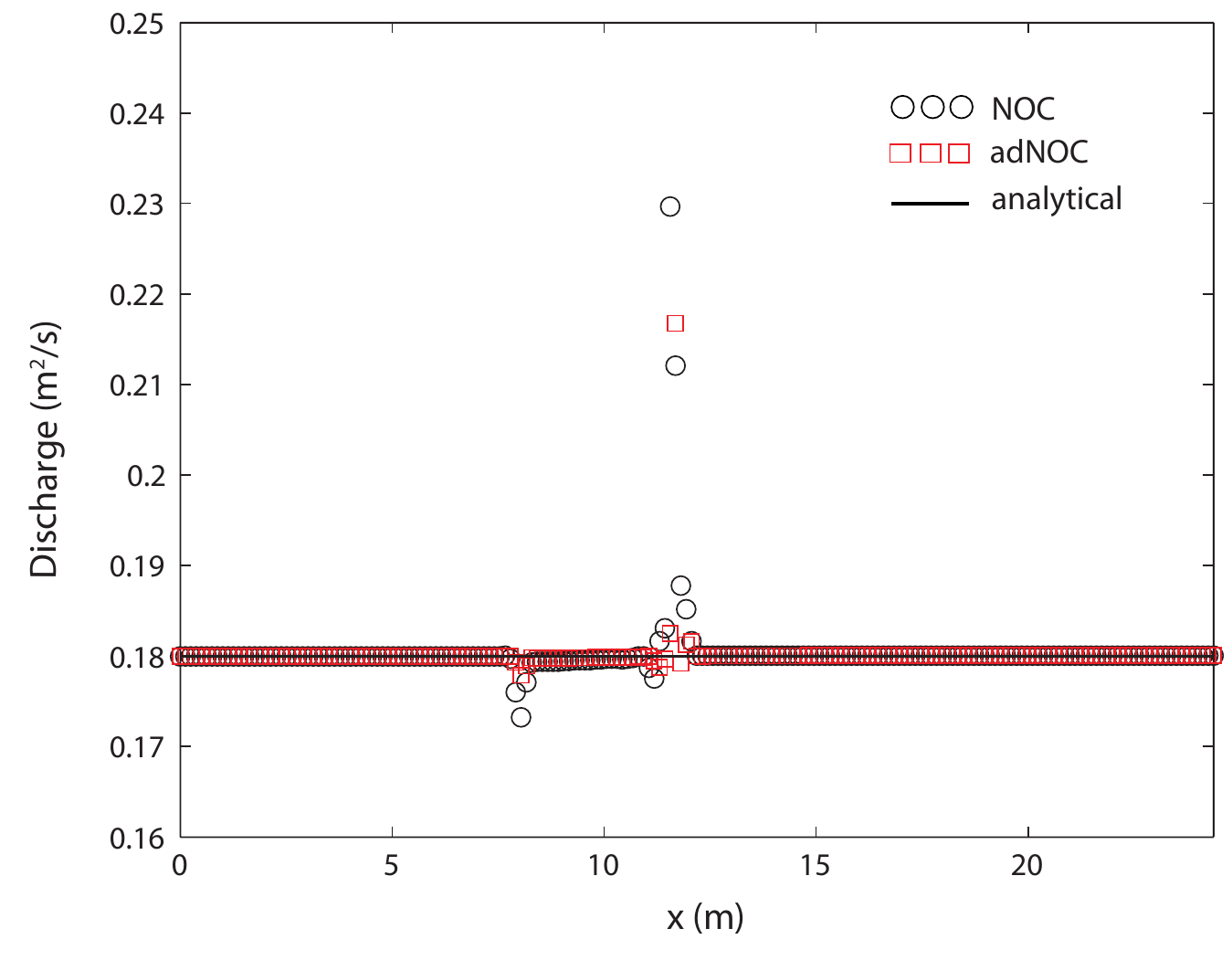}
\end{subfigure}

\begin{subfigure}{.5\textwidth}
  \centering
  \includegraphics[width=\linewidth]{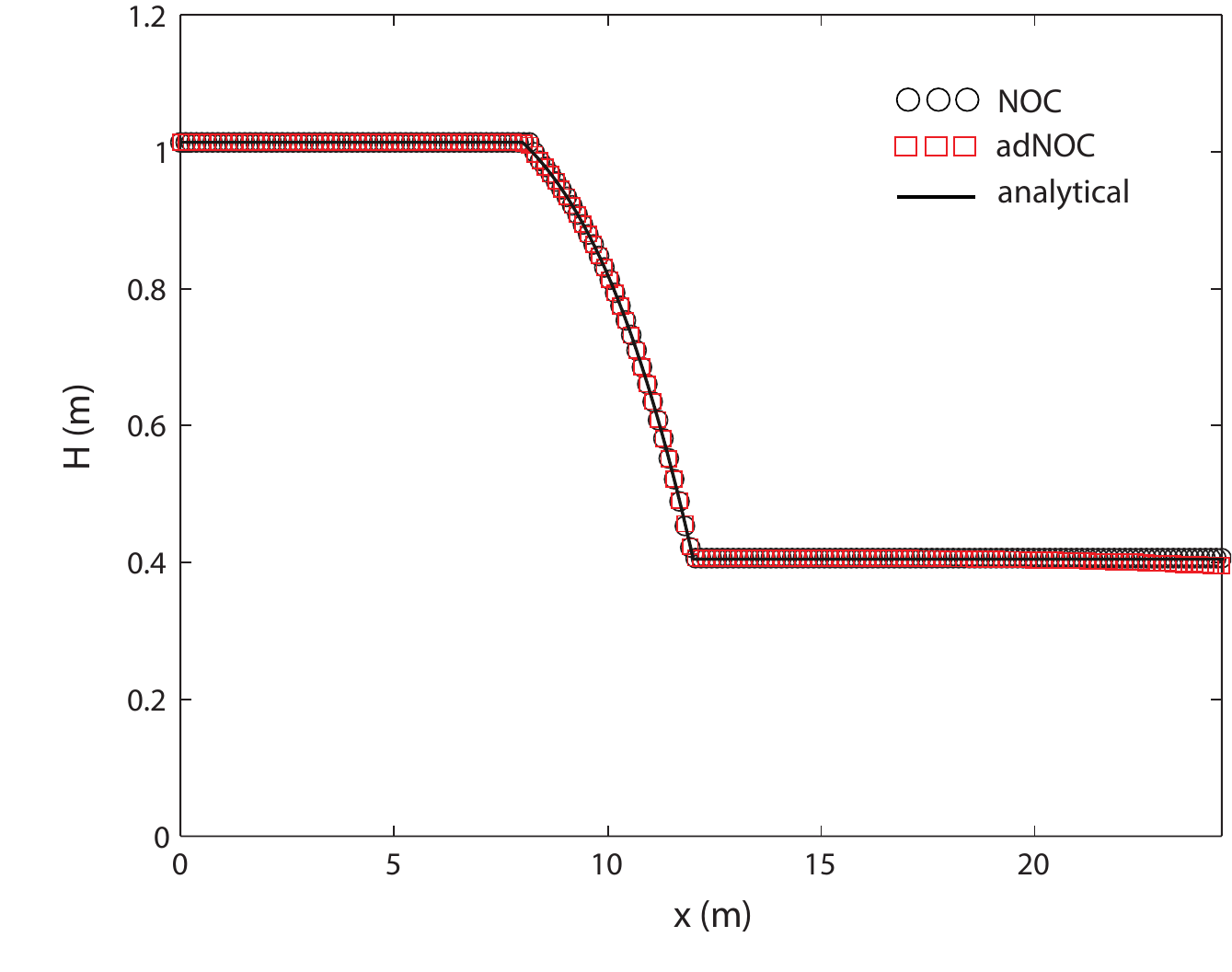}
\end{subfigure}%
\begin{subfigure}{.5\textwidth}
  \centering
  \includegraphics[width=\linewidth]{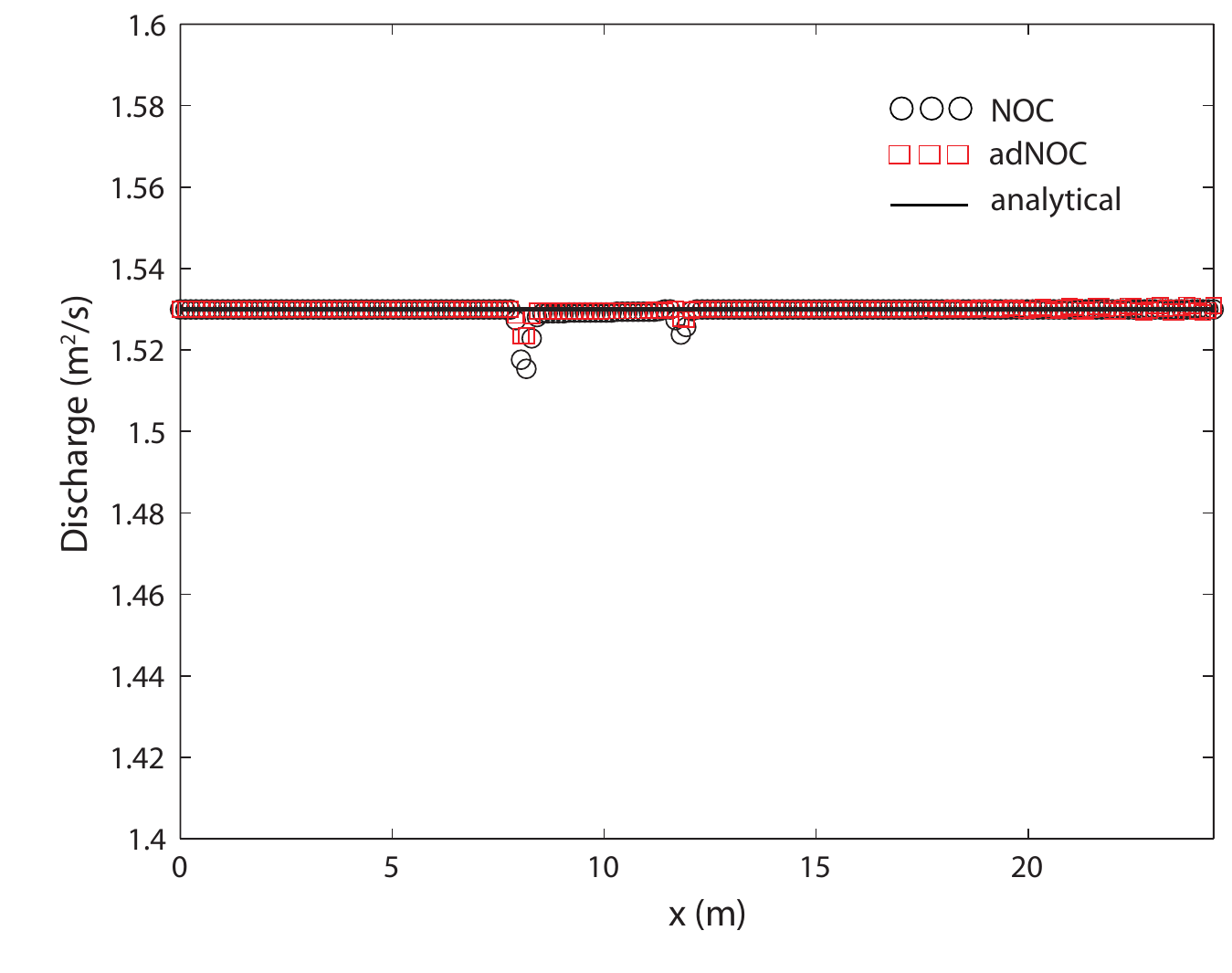}
\end{subfigure}
\caption{The water height $H$ and the Discharge $q$ at  steady state
calculated with the \textit{NOC} and \textit{adNOC} schemes. Test case (1) is shown at the top,
test case (2) in the middle and test case (3) is shown at the bottom.}
\label{fig:bump}
\end{figure}

The numerical experiment is performed by dividing the 25 m long domain into 200
cells. A courant number of 0.4 is used to calculate the time step while an
$\varepsilon$ of 0.6, 0.75 and 0.95 is used for the test case (1), (2) and (3) respectively. Fig \ref{fig:bump} shows the analytical solution as well as
the numerical results calculated with the \textit{NOC} and \textit{adNOC} schemes. Results
shows a good agreement between the analytical and numerical results. The small
errors appearing in the discharge are also suffered by other published schemes (e.g.
\cite{Canestrelli2010291,audusse2004fast,xing2006new}). It can be seen that the
\textit{adNOC} scheme shows better agreement with the analytical solution than
the \textit{NOC} scheme. This is especially true in test case (1), which does not have a sharp
discontinuity allowing a high value of $\varepsilon$ (0.95) to be used. This  shows
that the error reduces as higher values of $\varepsilon$ are used.

\subsection{Oscillatory flow in a parabolic bowl}
This popular test case was designed to demonstrate the ability of numerical schemes to
accurately and robustly capture  dynamic wet-dry boundaries (e.g.
\cite{kesserwani2010discontinuous,liang2009numerical,liang2009adaptive,kesserwani2010well}).
The test case consists of flow inside a convex domain with a parabolic
profile. The forces generated by the parabolic topography results in  flow
that oscillates from one side of the parabola to the other. An analytical
solution of the problem was provided by Thacker \cite{thacker1981some}, which was later  extended to include the friction slope by Sampson et al.
\cite{sampson2006moving}. The parabolic profile of the topography is given by:
\[
z(x)=h_{0}\left(\frac{x}{a}\right)^{2},
\]
where $h_0$ and $a$ are constants. The analytical solution for a frictionless
bottom is given by:
\[
\zeta(x,t)=h_{0}-\frac{a^{2}B^{2}}{8g^{2}h_{0}}(s^{2}cos(2st))-\frac{B^{2}}{4g}-\frac{Bs\;
cos(st)}{g}x,
\]
where $B$ is a constant and $s=\sqrt{8gh_0/4a^2}$. The location of the moving
wet-dry boundary can be calculated by:
\[
x=\frac{a^{2}}{2gh_{0}}\left(Bs\; cos(st)\right)\pm a
\]

\begin{figure}
	\centering
	\includegraphics[scale=0.7]{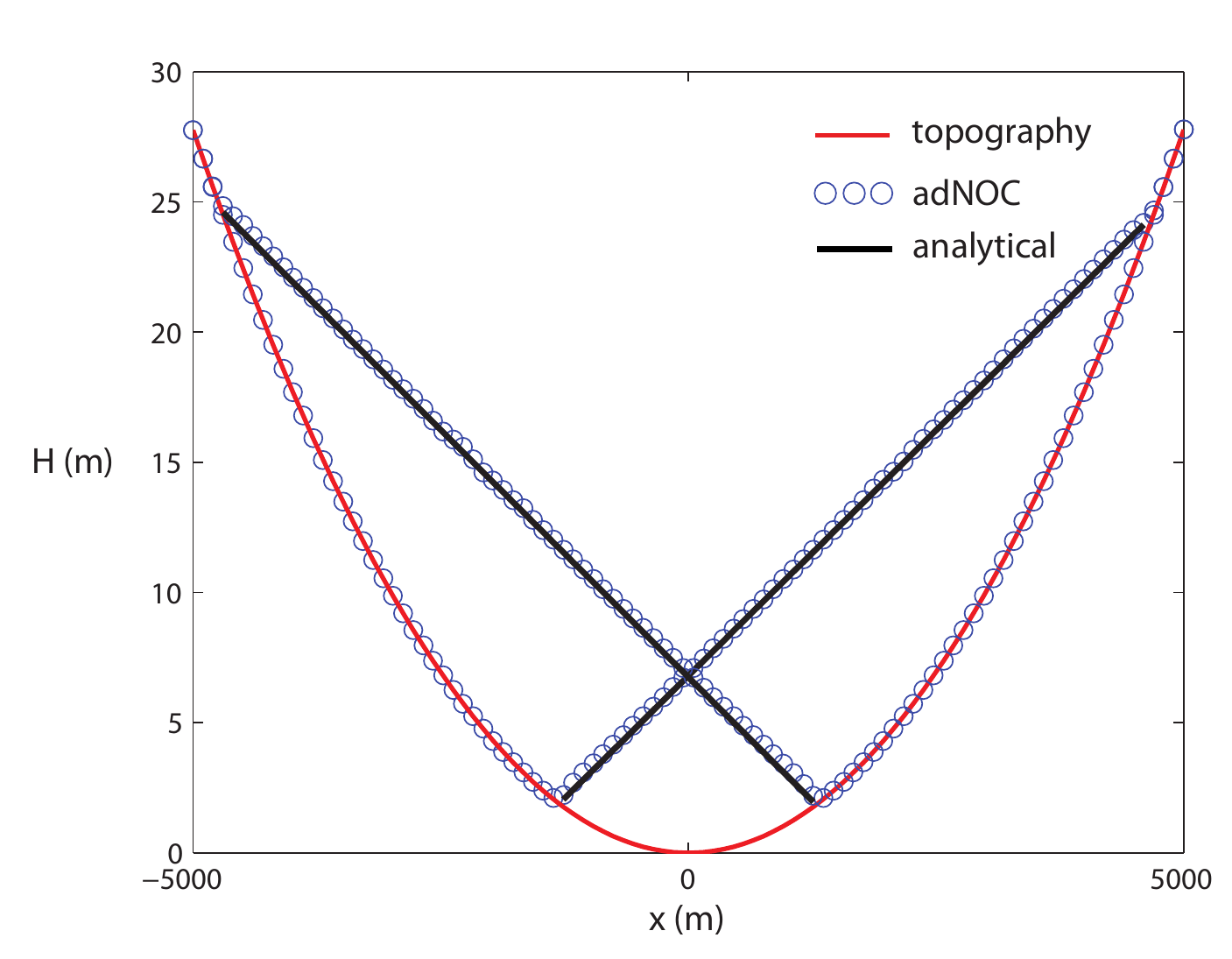}
		\caption{Flow over a parabola bed topography. The analytical and the numerical
		solution after $1\frac{1}{2}$ and 2 time periods of the oscillation.}
	\label{fig:parabola}
\end{figure}

We have performed  numerical simulations on a 10,000 m long domain, discretised  with 100 cells. The parameter values used to calculate the analytical
solution $\zeta$ are $h_0=10$, $a=3000$ and $B=8$. The simulation is started
with the following initial conditions:
\[
 \begin{cases} H(x)=\zeta(x,0)  &\mbox{if } \zeta(0)>z  \\
 u=0. \end{cases}
\]
 The time step is calculated using a Courant number of $C_n=0.4$ and an
 $\varepsilon$ of 0.6 and 0.2  is used for the continuity and momentum equations, respectively. Fig.
\ref{fig:parabola} shows the numerical and the analytical solution after 1.5
and 2 time periods of the oscillations. It can be seen that the numerical
results have an excellent agreement with the analytical solution.

\section{Perturbation of a lake at rest with discontinuous bottom topography}
\begin{figure}
	\centering
	\includegraphics[scale=0.7]{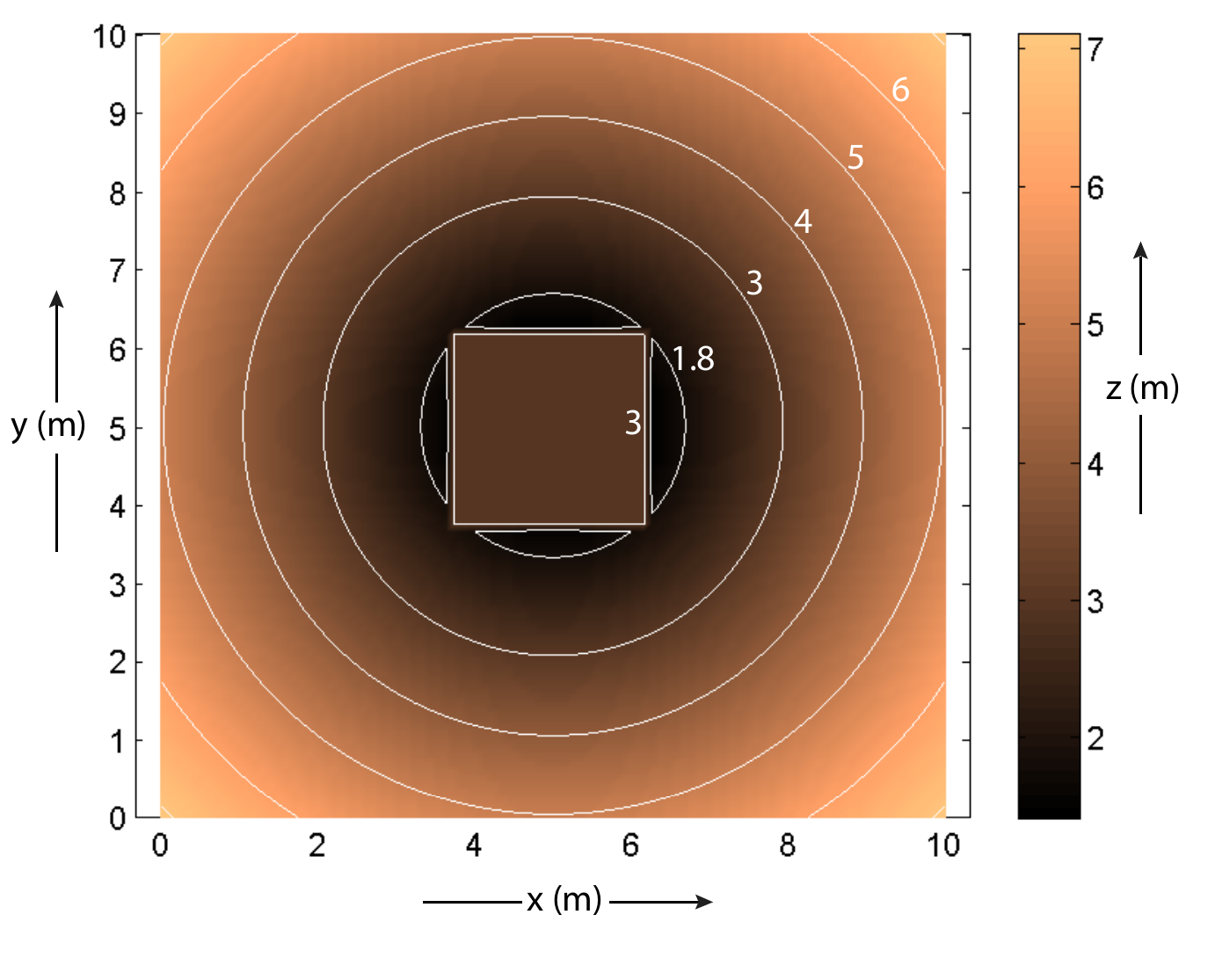}
		\caption{Bed topography with a sharp step in the center.}
	\label{fig:bed}
\end{figure}

This numerical experiment designed for the first time here aims to demonstrate the ability of the \textit{adNOC}
scheme to robustly handle a number of scenarios commonly encountered in
nature. The test is challenging as it comprises simultaneously of three
scenarios which are difficult for numerical schemes to simulate. The first is the
'lake at rest' condition, the second is the accurate resolution of flow over discontinuous
bottom topography and the third is the robust handling of wet-dry fronts.
The test is performed in two dimensions using the two dimensional \textit{adNOC}
scheme (see \cite{Zia2015anti}). The setup consists of a lake initially at rest,
which is given a small perturbation. The time evolution of the water level
is tracked until the lake is once again at rest  (bottom
friction is included). The bottom topography is in the form of a bowl with a sharp square
step of size 1.5 $m$ in the middle, which aims at testing how well the scheme is
able to uphold the so called "well balanced'' property. The topography in the 10x10 m
domain is given by (see Fig \ref{fig:bed}):
\[
 z(x,y)=\begin{cases} 3\;m &\mbox{if } \;\;3.75\;m\leq x \leq 6.25\;m\;\;
 \text{and}\;\; 3.75\;m\leq y \leq 6.25\;m\\
 \sqrt{(x-5)^2+(y-5)^2+0.4}\;m &\mbox{} \text{otherwise}
\end{cases}
\]

 The water level is set
to 4 $m$ with an additional perturbation in the center. The initial condition for the
water height $H$ is given by (see Fig. \ref{fig:lake}):
\[
 H(x,y)=\begin{cases} 4.6-(x-5)^2+(y-5)^2\;m &\mbox{if }
 \;\;(x-5)^2+(y-5)^2\leq 0.6\;m\\
 4\;m &\mbox{if } \;\;z<4\;m\\
 z &\mbox{} \text{otherwise}
\end{cases}
\]

The numerical simulation is carried out with 100 cells in each direction. We have used $\varepsilon$
values of 1 and 0.9 for the continuity equation and the momentum conservation
equations, respectively. These relatively high values are necessary for the scheme to
be fully "well-balanced" but can make the scheme unstable . We have avoided
instability by enforcing a sufficiently small courant number of 0.01.
Also, since the water depth appears in the denominator of the friction slope
formulation, a threshold of 0.003 $m$ for the water depth is
introduced below which the solution is ignored and the calculations are not
performed to avoid singularities.

The results of the simulation are shown in Fig. \ref{fig:lake}. The perturbation
in the middle causes a wave to move radially outwards, which is
reflected back after it hits the shoreline. It can be seen that the
discontinuity at the bottom is effectively handled  and a 'numerical storm' is
avoided. Although the test case is performed using a small courant number, it
demonstrates the flexibility of the scheme. By using a high $\varepsilon$ (with
a possible cost of small time step), numerically challenging problems such as
this test case can be successfully modeled.
\begin{figure}
\centering
\begin{subfigure}{.5\textwidth}
  \centering
  \includegraphics[width=\linewidth]{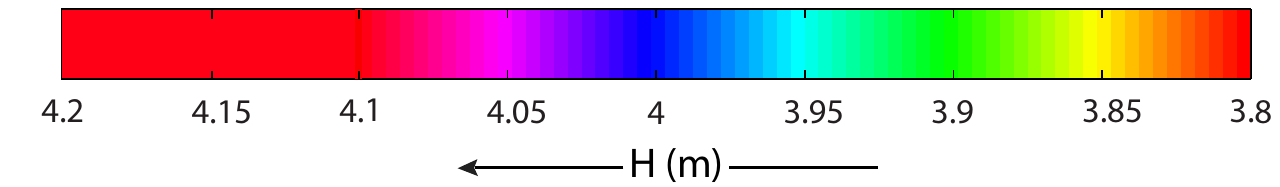}
\end{subfigure}%
\begin{subfigure}{.5\textwidth}
  \centering
  \includegraphics[width=\linewidth]{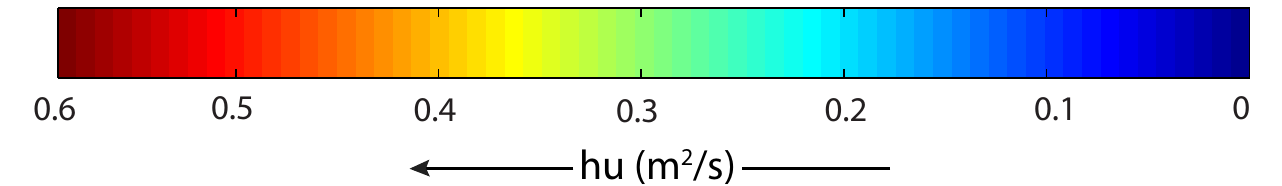}
\end{subfigure}

\begin{subfigure}{.5\textwidth}
  \centering
  \includegraphics[width=\linewidth]{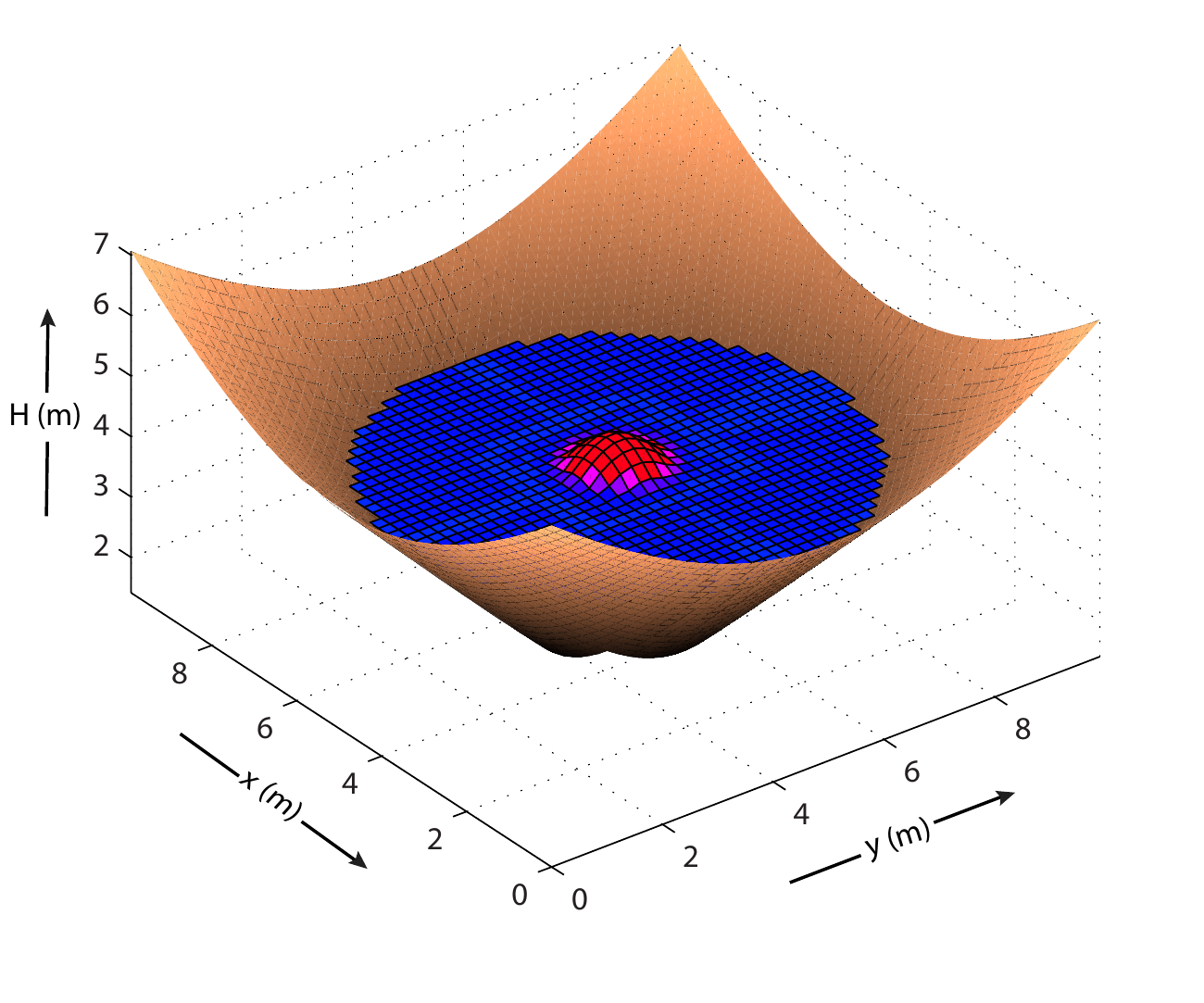}
  \captionsetup{labelformat=empty}
  \caption{water level at 0 sec.}
\end{subfigure}%
\begin{subfigure}{.5\textwidth}
  \centering
  \includegraphics[width=\linewidth]{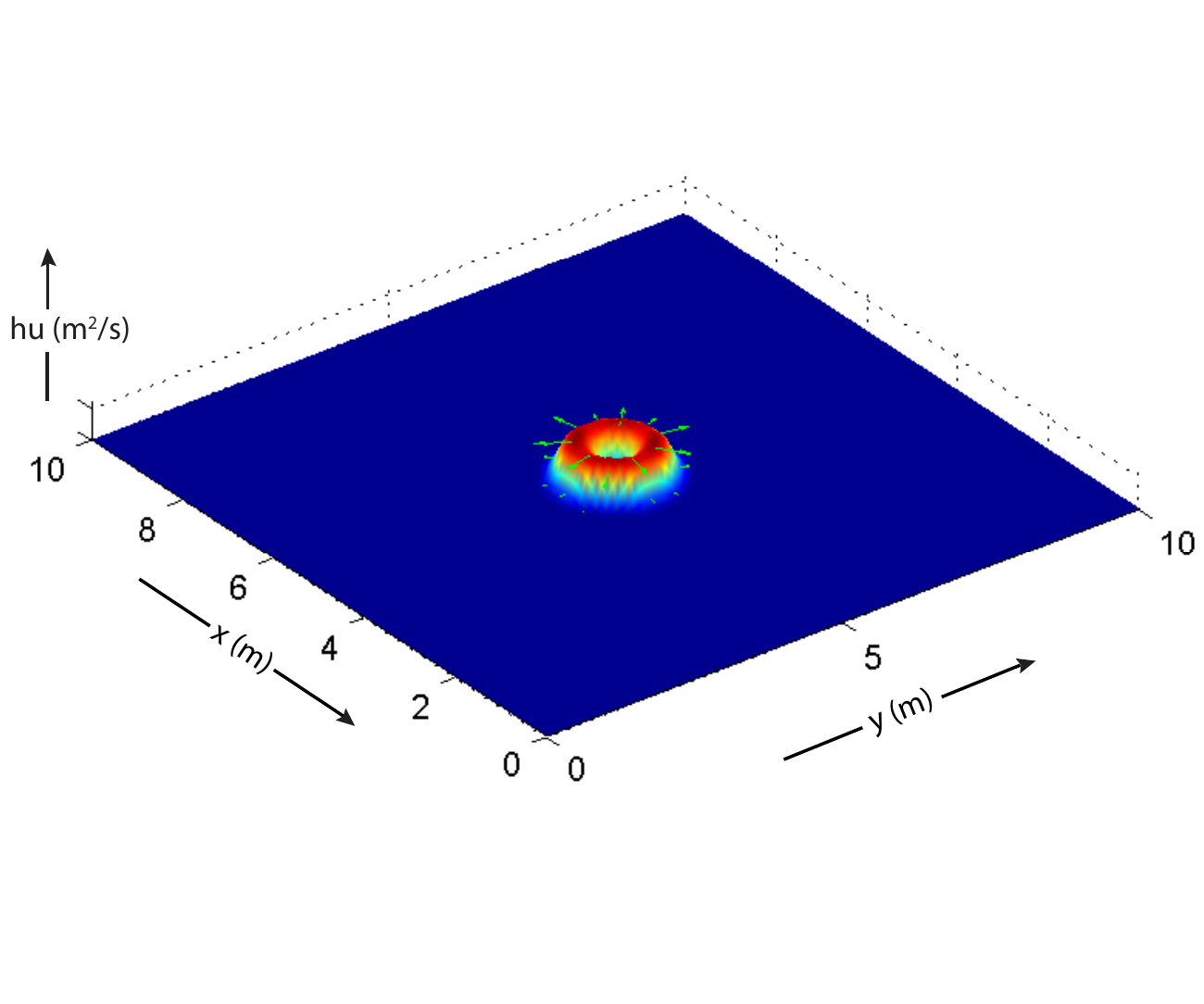}
  \captionsetup{labelformat=empty}
  \caption{discharge at 0 sec.}
\end{subfigure}

\begin{subfigure}{.5\textwidth}
  \centering
  \includegraphics[width=\linewidth]{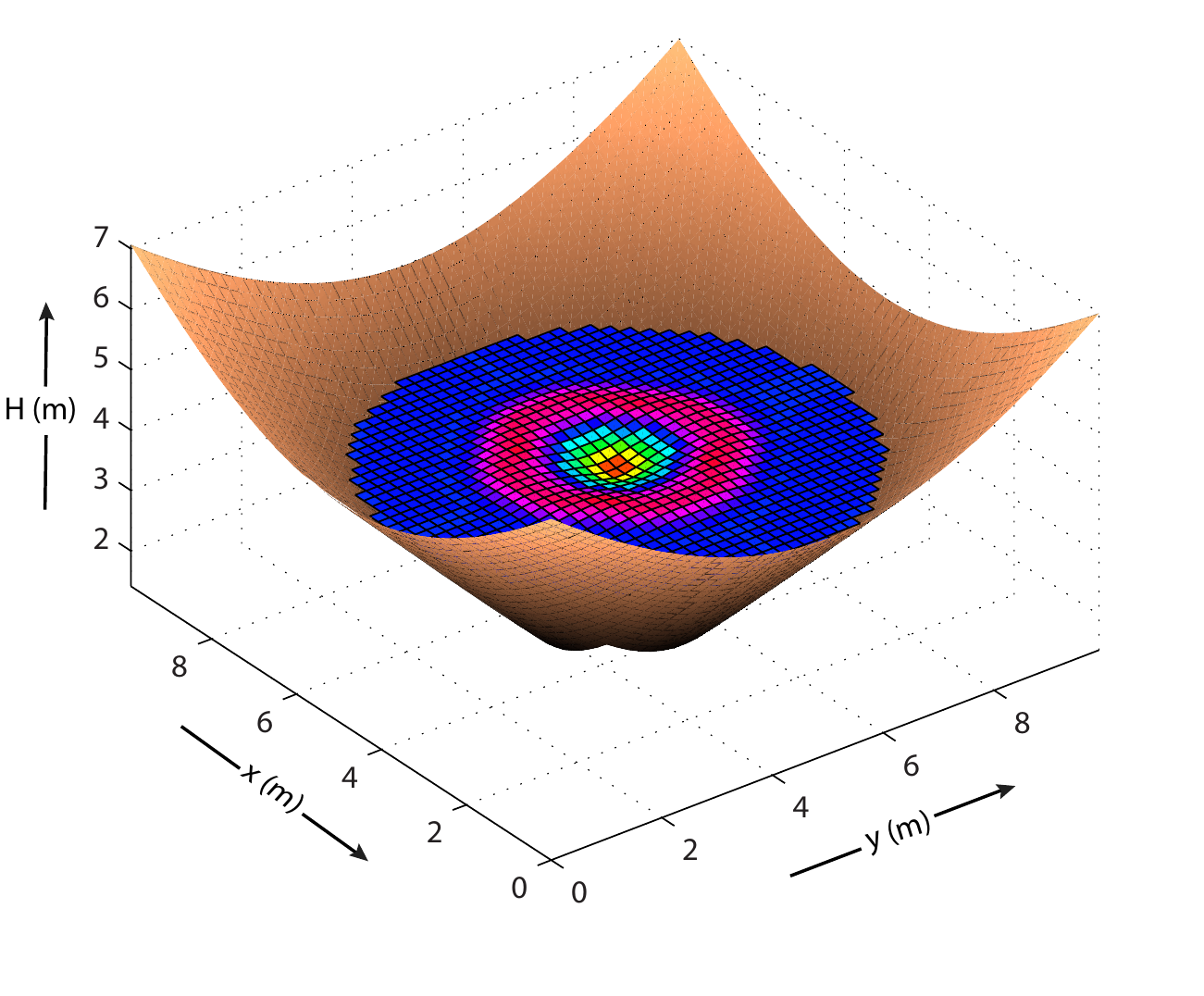}
  \captionsetup{labelformat=empty}
  \caption{water level at 0.35 sec.}
\end{subfigure}%
\begin{subfigure}{.5\textwidth}
  \centering
  \includegraphics[width=\linewidth]{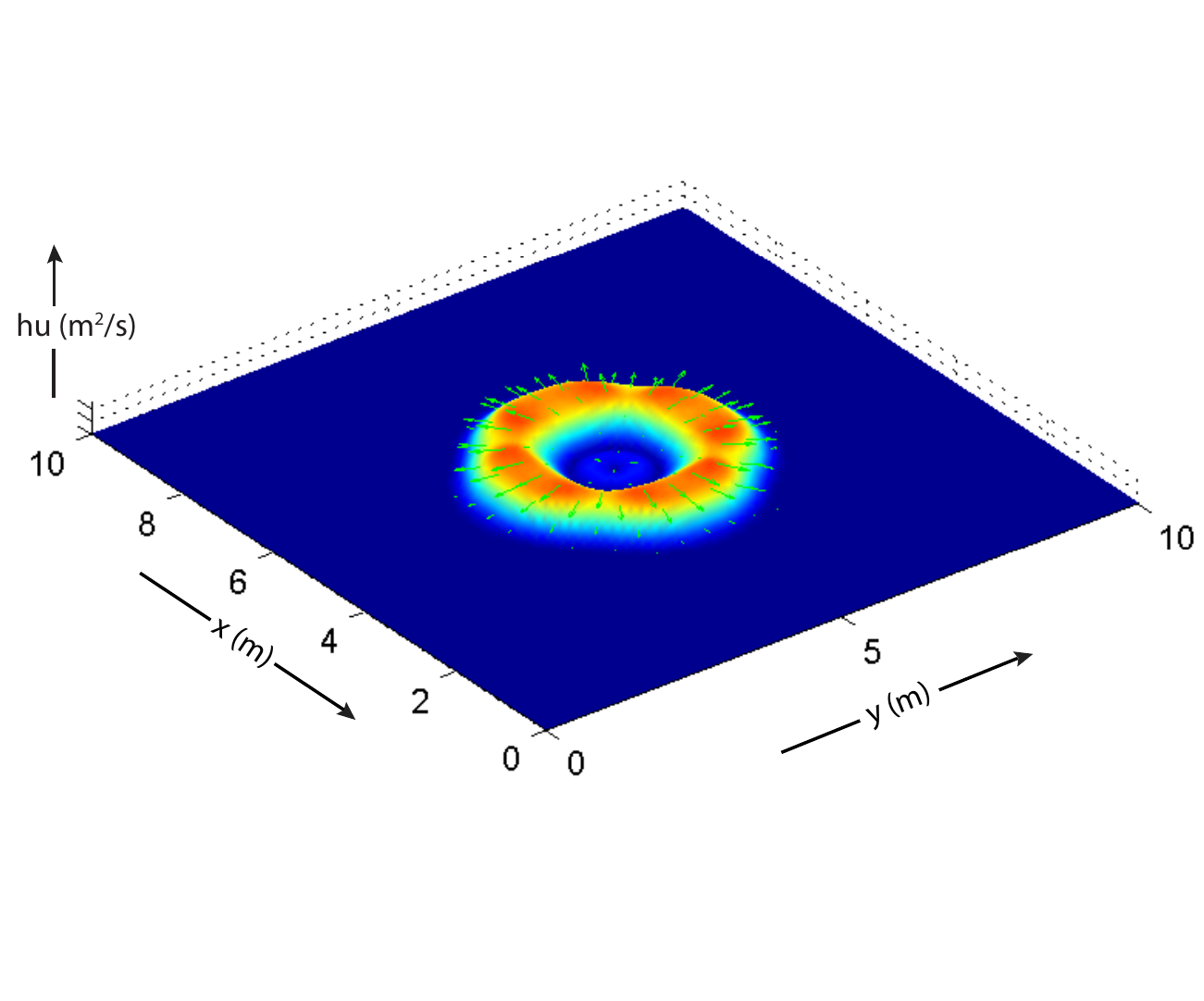}
  \captionsetup{labelformat=empty}
  \caption{discharge at 0.35 sec.}
\end{subfigure}

\begin{subfigure}{.5\textwidth}
  \centering
  \includegraphics[width=\linewidth]{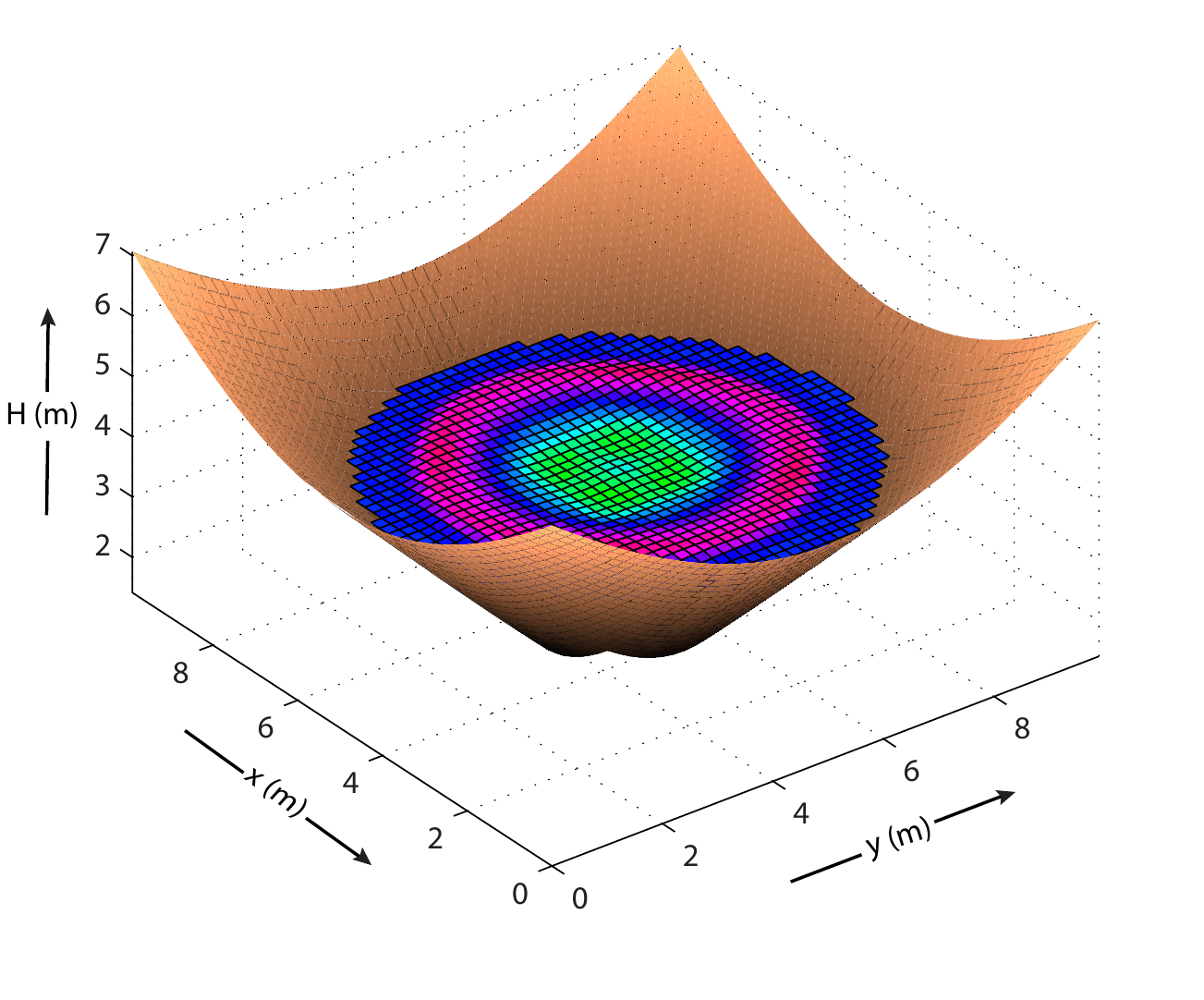}
  \captionsetup{labelformat=empty}
  \caption{water level at 0.60 sec.}
\end{subfigure}%
\begin{subfigure}{.5\textwidth}
  \centering
  \includegraphics[width=\linewidth]{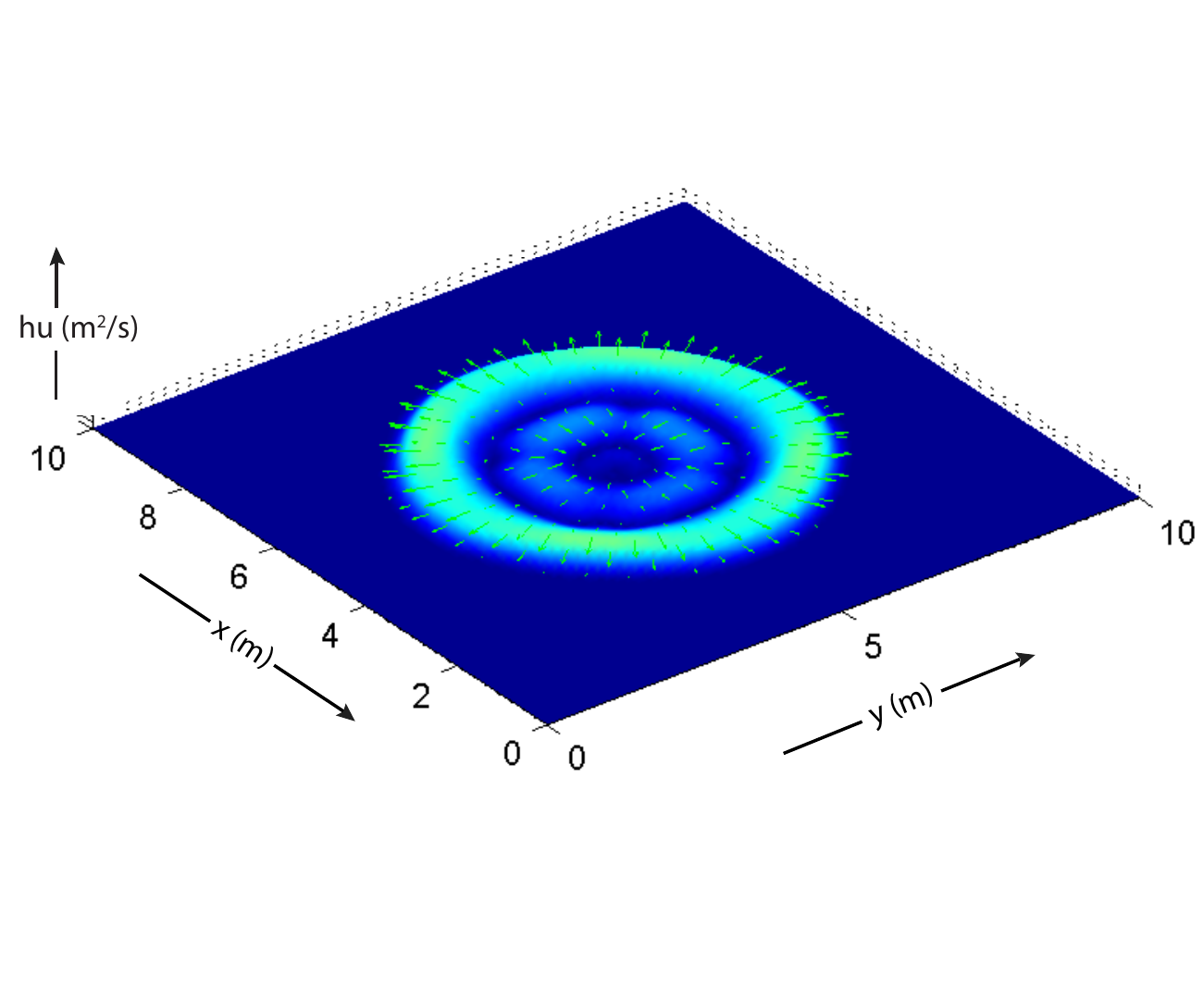}
  \captionsetup{labelformat=empty}
  \caption{discharge at 0.60 sec.}
\end{subfigure}
\end{figure}

\begin{figure}
\ContinuedFloat
\centering
\begin{subfigure}{.5\textwidth}
  \centering
  \includegraphics[width=\linewidth]{bar_lake.pdf}
\end{subfigure}%
\begin{subfigure}{.5\textwidth}
  \centering
  \includegraphics[width=\linewidth]{bar_mmt.pdf}
\end{subfigure}

\begin{subfigure}{.5\textwidth}
  \centering
  \includegraphics[width=\linewidth]{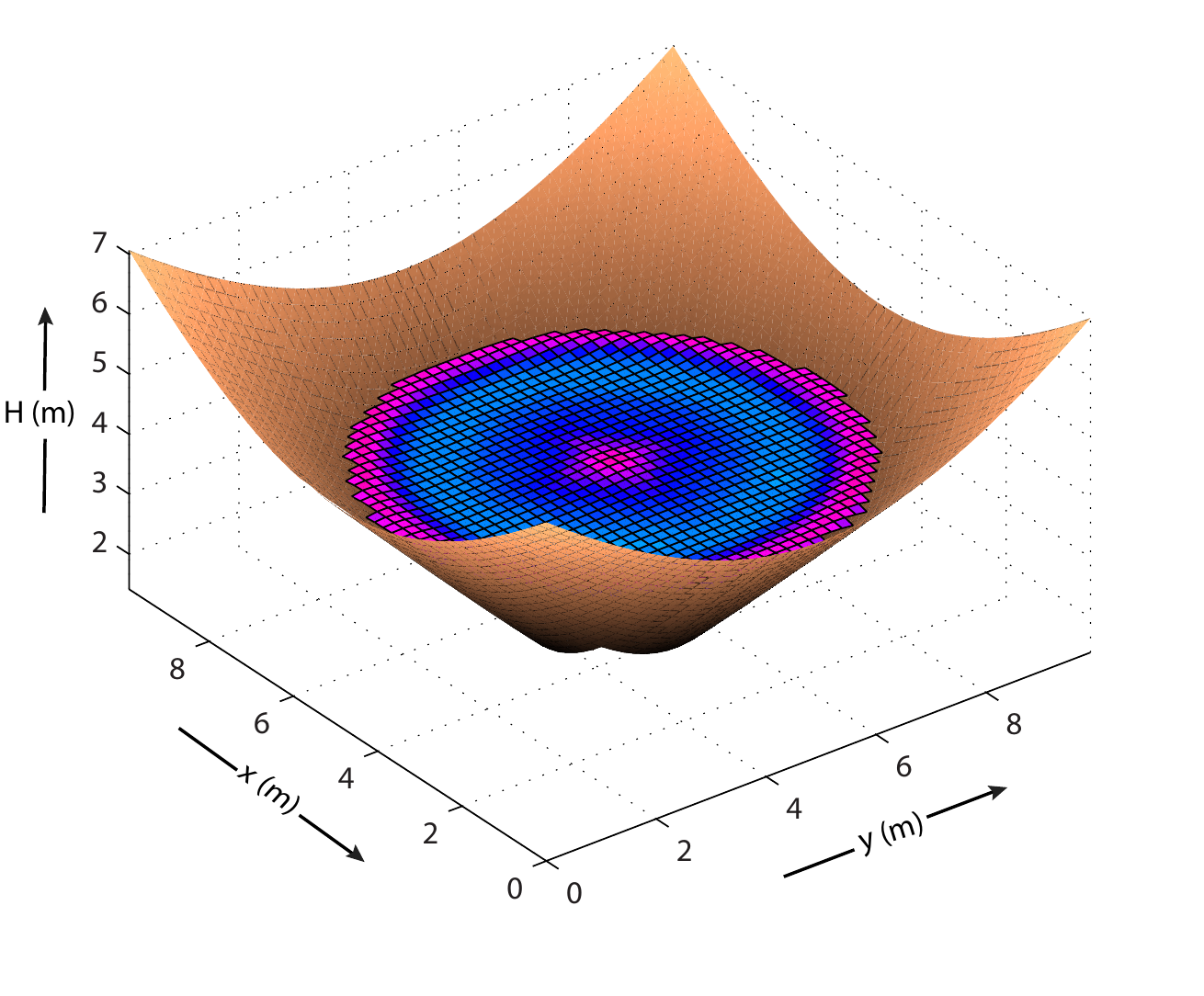}
  \captionsetup{labelformat=empty}
  \caption{water level at 1.00 sec.}
\end{subfigure}%
\begin{subfigure}{.5\textwidth}
  \centering
  \includegraphics[width=\linewidth]{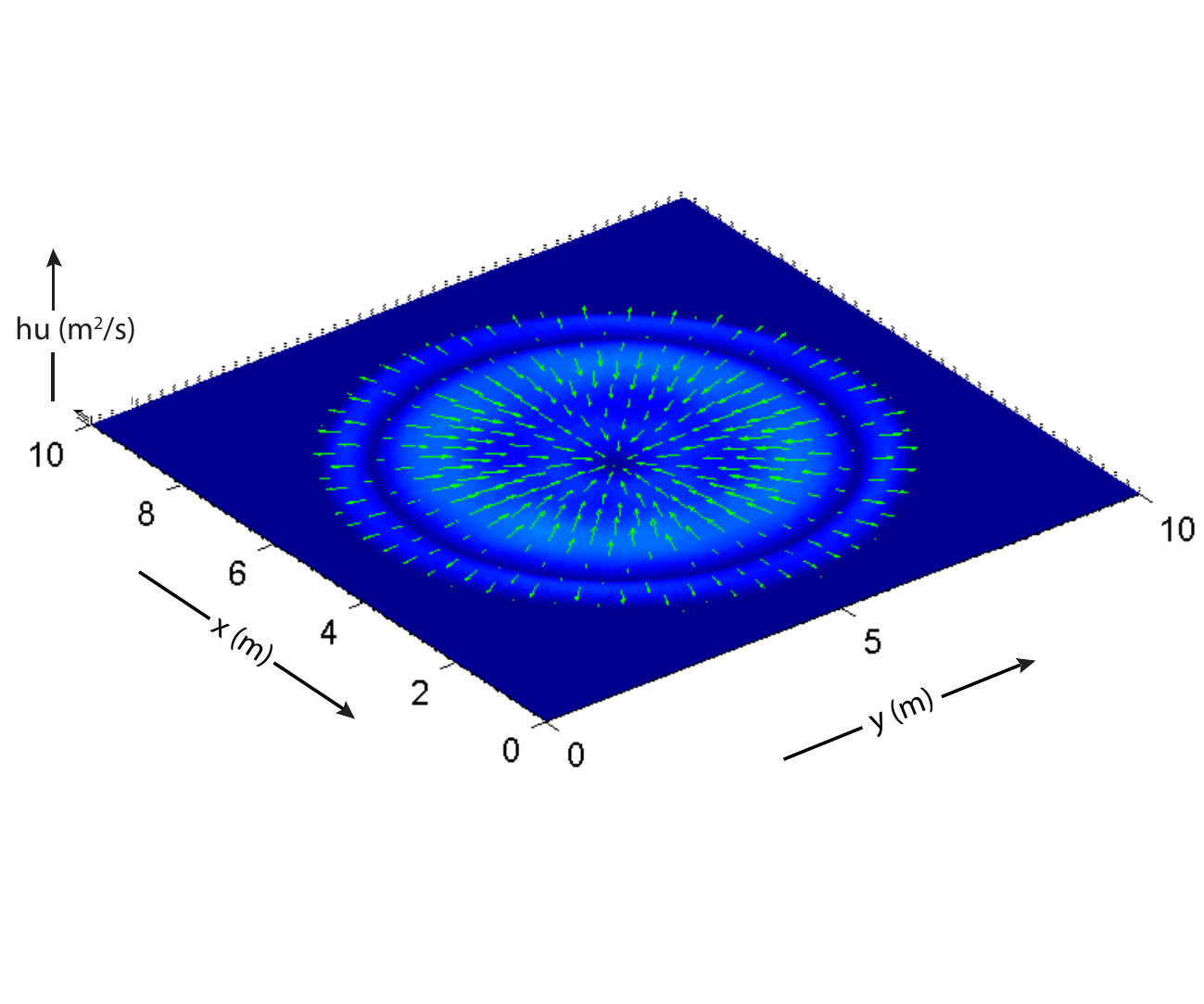}
  \captionsetup{labelformat=empty}
  \caption{discharge at 1.0 sec.}
\end{subfigure}

\begin{subfigure}{.5\textwidth}
  \centering
  \includegraphics[width=\linewidth]{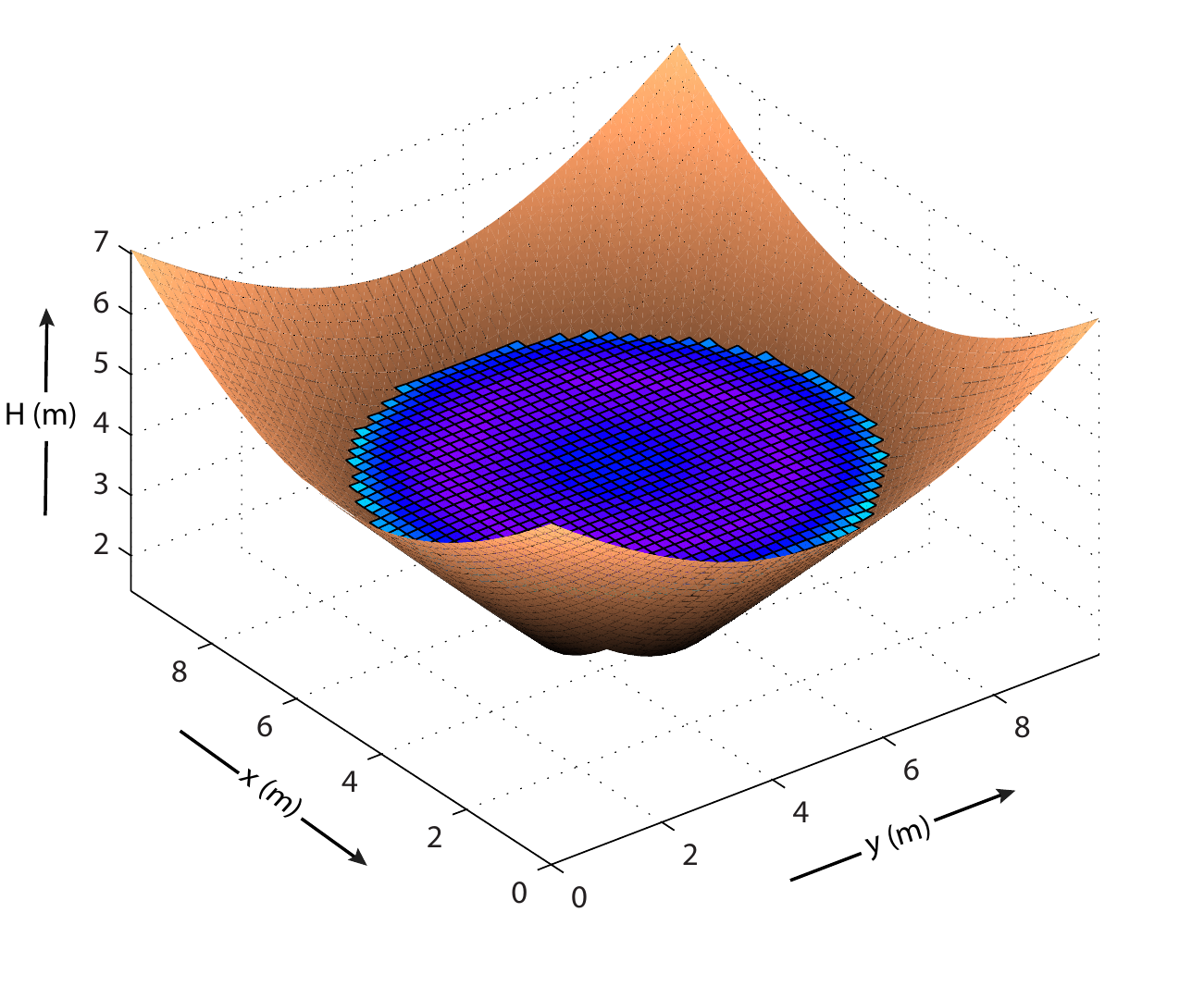}
  \captionsetup{labelformat=empty}
  \caption{water level at 1.7 sec.}
\end{subfigure}%
\begin{subfigure}{.5\textwidth}
  \centering
  \includegraphics[width=\linewidth]{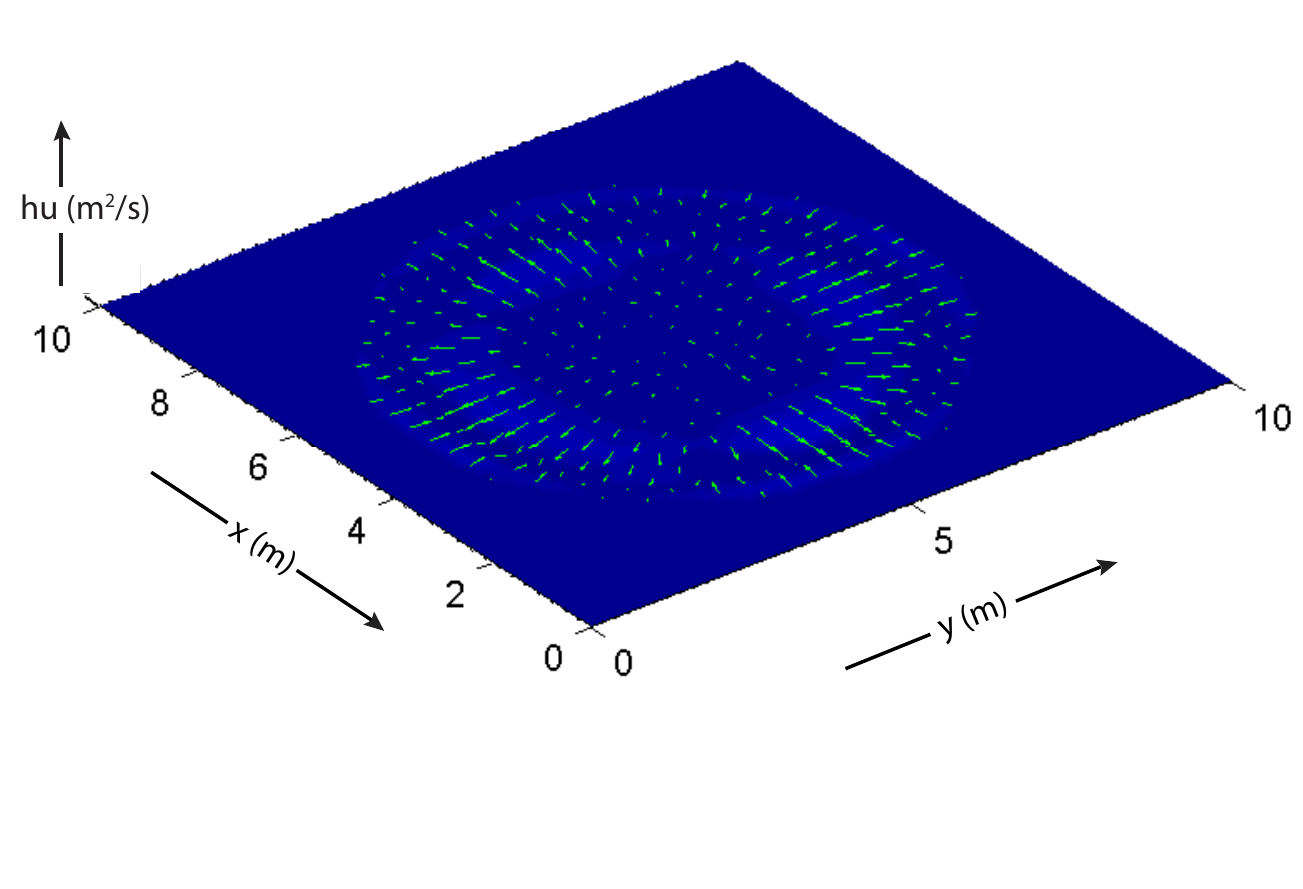}
  \captionsetup{labelformat=empty}
  \caption{discharge at 1.7 sec.}
\end{subfigure}

\begin{subfigure}{.5\textwidth}
  \centering
  \includegraphics[width=\linewidth]{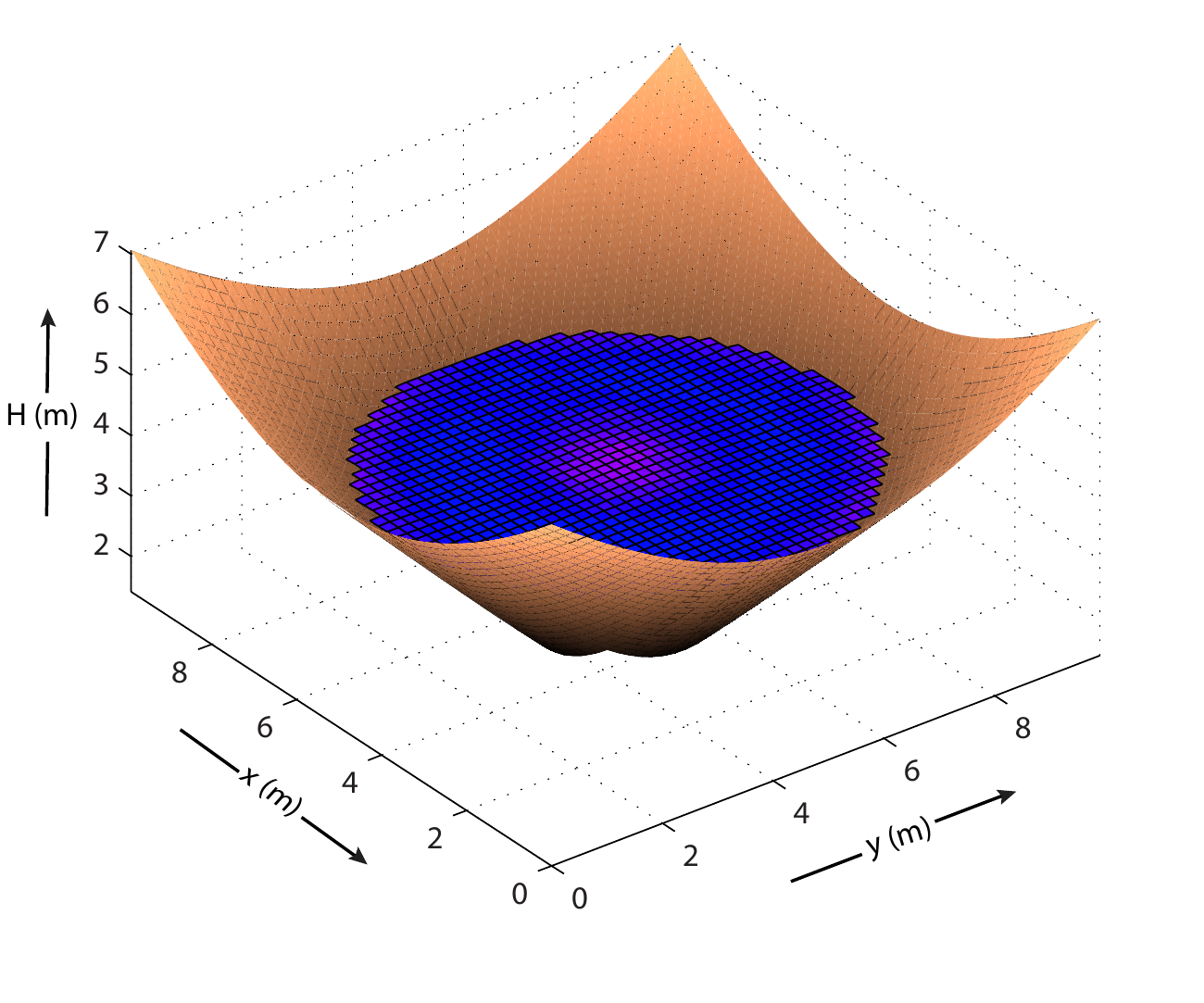}
  \captionsetup{labelformat=empty}
  \caption{water level at 2.25 sec.}
\end{subfigure}%
\begin{subfigure}{.5\textwidth}
  \centering
  \includegraphics[width=\linewidth]{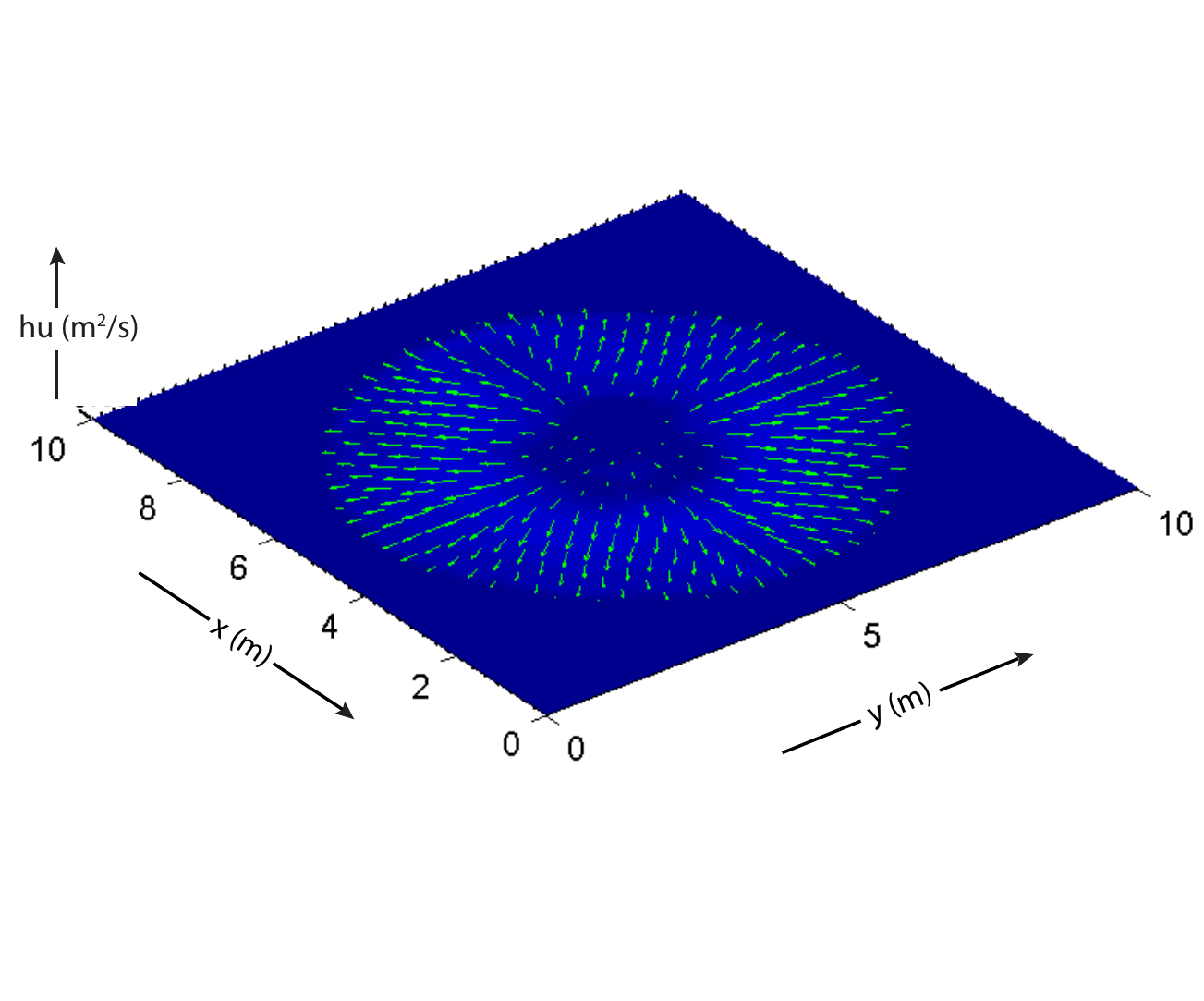}
  \captionsetup{labelformat=empty}
  \caption{discharge at 2.25 sec.}
\end{subfigure}
\end{figure}

\begin{figure}
\ContinuedFloat
\centering
\begin{subfigure}{.5\textwidth}
  \centering
  \includegraphics[width=\linewidth]{bar_lake.pdf}
\end{subfigure}%
\begin{subfigure}{.5\textwidth}
  \centering
  \includegraphics[width=\linewidth]{bar_mmt.pdf}
\end{subfigure}

\begin{subfigure}{.5\textwidth}
  \centering
  \includegraphics[width=\linewidth]{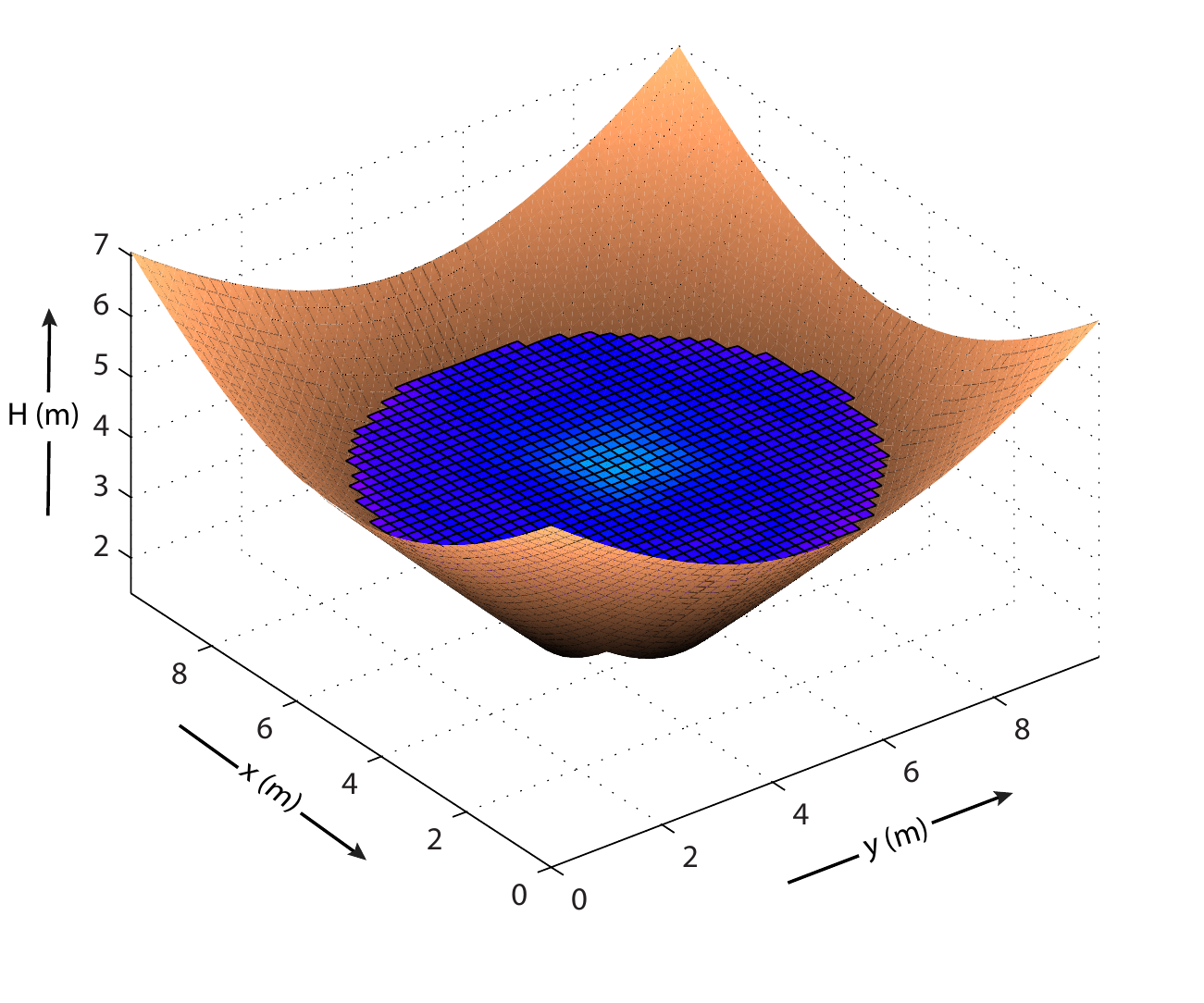}
  \captionsetup{labelformat=empty}
  \caption{water level at 2.95 sec.}
\end{subfigure}%
\begin{subfigure}{.5\textwidth}
  \centering
  \includegraphics[width=\linewidth]{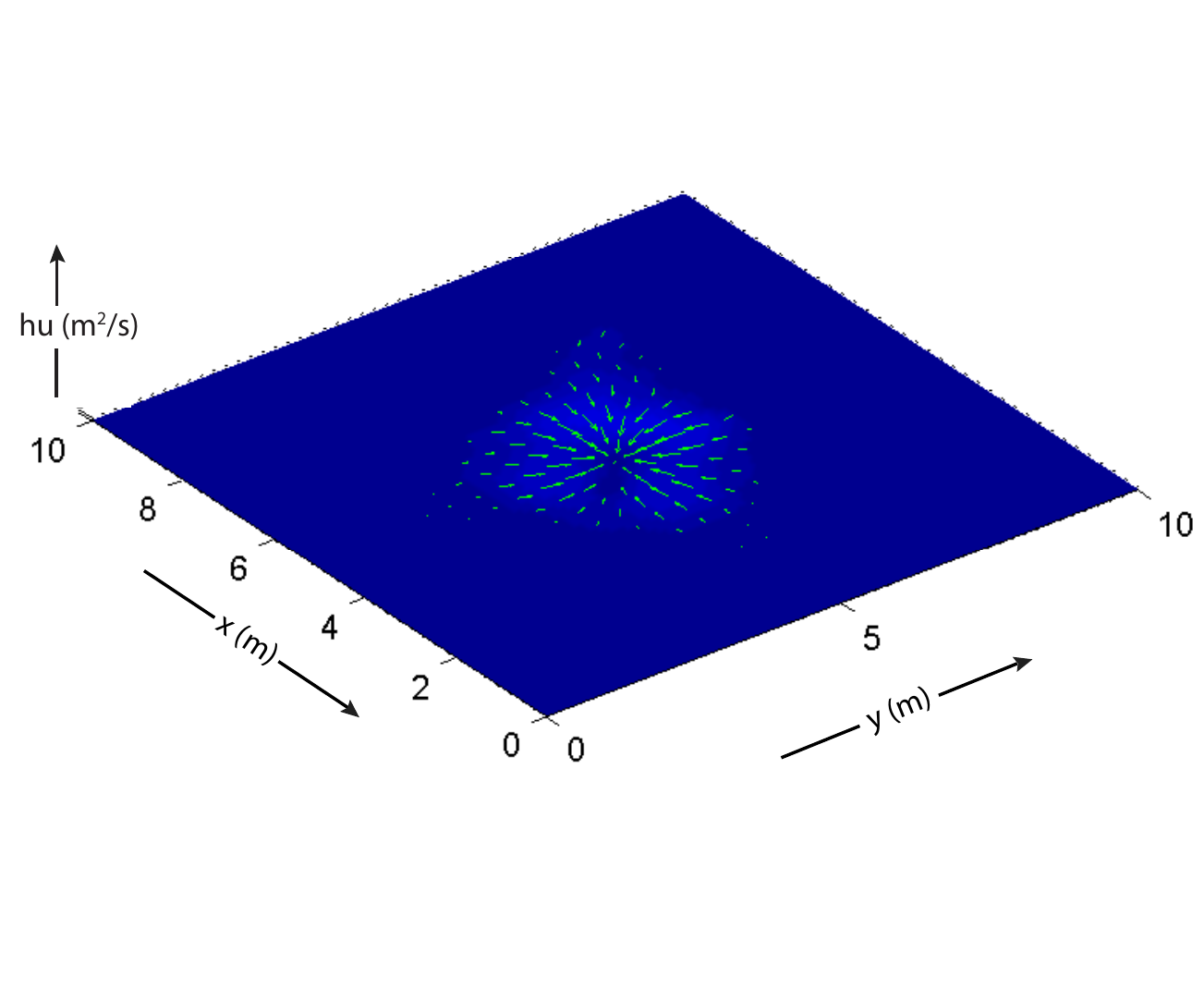}
  \captionsetup{labelformat=empty}
  \caption{discharge at 2.95 sec.}
\end{subfigure}

\begin{subfigure}{.5\textwidth}
  \centering
  \includegraphics[width=\linewidth]{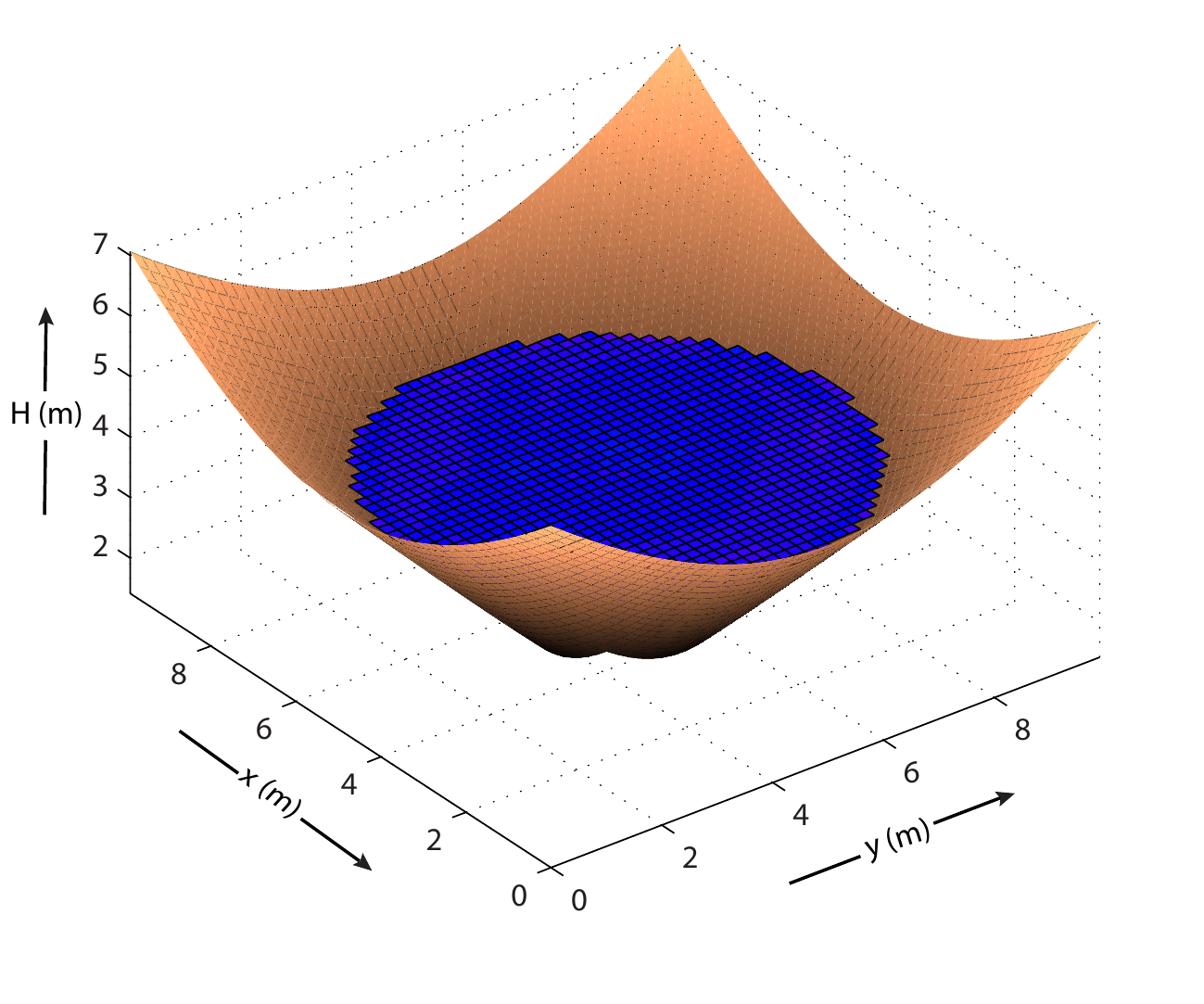}
  \captionsetup{labelformat=empty}
  \caption{water level at steady state.}
\end{subfigure}%
\begin{subfigure}{.5\textwidth}
  \centering
  \includegraphics[width=\linewidth]{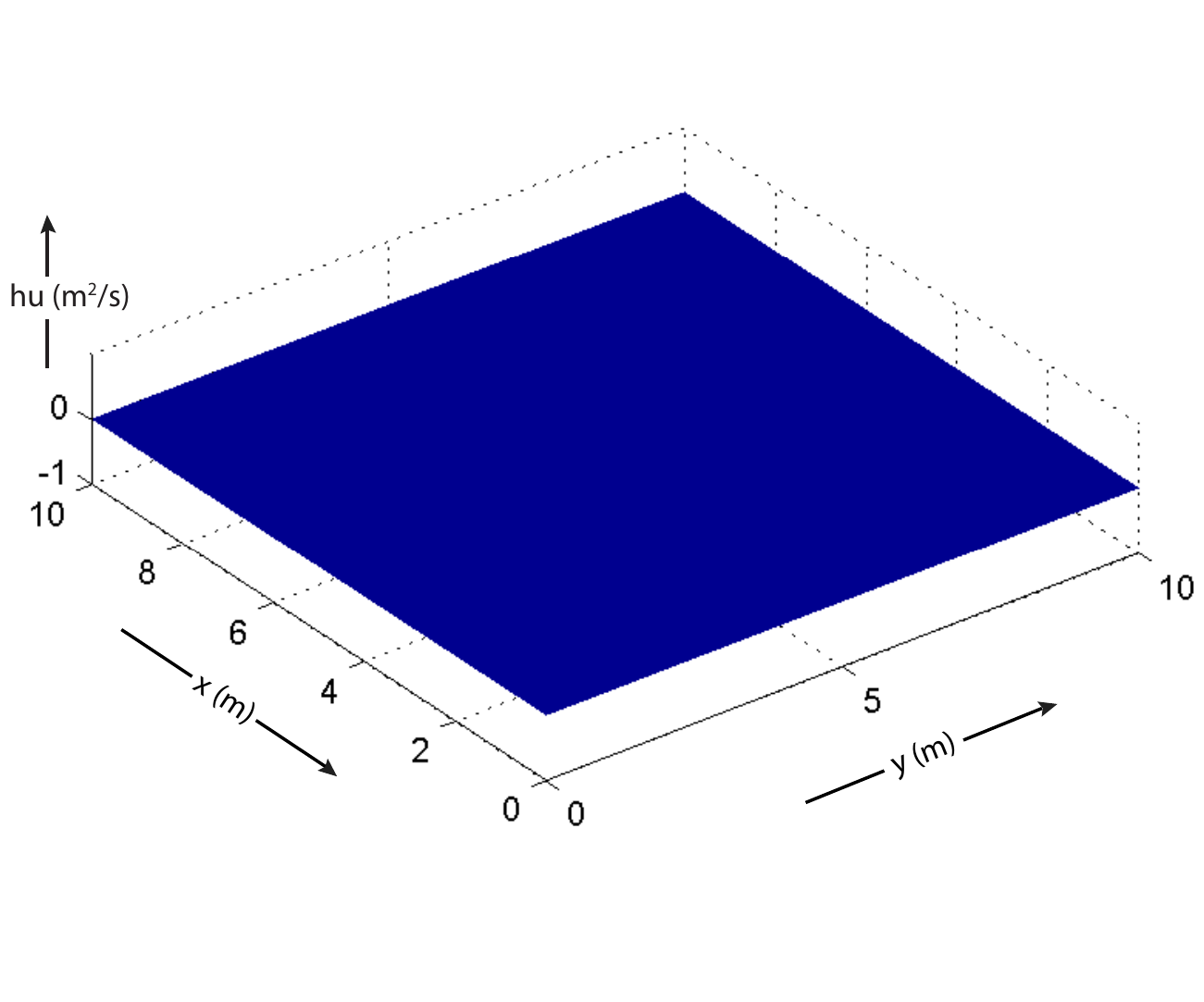}
  \captionsetup{labelformat=empty}
  \caption{discharge at steady state.}
\end{subfigure}

\caption{Water height and discharge for lake test case. The solution is calculated with the \textit{adNOC} scheme. The
arrows in the discharge figures signify the direction of the discharge while the
color signifies the amplitude.}
\label{fig:lake}
\end{figure}

\section{Conclusion}
We have presented  a simple, fast and robust central scheme to
solve the shallow water equations  over discontinuous bottom topography. The scheme upholds the
well-balanced C-property by exactly balancing the flux
gradient and source terms. The well-balancing is done by more precise
calculation of the flux gradient and by reducing the discretization error by using
the same finite difference operators for both the flux gradient and the source
terms. The scheme is  shown to be positivity preservating if a sufficiently
small time step is used. The numerical performance of the scheme is validated with
three test cases. First, the scheme is tested for correct evaluation of the steady
state solutions in the case of quiescent flow over a bump. Second, the scheme is
tested for its robust performance in the presence of dynamic wet-dry boundaries in the case for flow over parabolic topography. Finally, the performance of the scheme is tested
with a two-dimensional case involving a sharp discontinuity at the bottom.
The final test is particularly challenging since it involves contemporaneous handling of wet-dry boundaries,
bed friction and discontinuous bottom topography. It is shown
that the scheme handles all three challenges effectively.
\newpage
\bibliography{Manuscript}
\end{document}